\documentclass[12pt]{iopart}
 
\usepackage{iopams}
\usepackage{amssymb}
\eqnobysec
\newtheorem{theorem}{Theorem}[section]
     \newtheorem{lemma}[theorem]{Lemma}
     
     \newtheorem{corollary}[theorem]{Corollary}
     \newenvironment{proof}[1][Proof.]{\begin{trivlist}
     \item[\hskip \labelsep {\bfseries #1}]}{\end{trivlist}}

     \newenvironment{remark}[1][Remark.]{\begin{trivlist}
     \item[\hskip \labelsep {\bfseries #1}]}{\end{trivlist}}

\begin{document}

\title[]{The Dirac propagator in the Kerr-Newman metric}

\author{D. Batic}
\address{Institute for Theoretical Physics, ETH Z\"{u}rich, CH-8093 Z\"{u}rich, Switzerland}
\ead{batic@itp.phys.ethz.ch}
\author{H. Schmid}
\address{UBH Software \& Engineering GmbH, D-92224 Amberg, Germany}
\ead{Harald.Schmid@UBH.d}

\begin{abstract}
We give an alternative proof of the completeness of the Chandrasekhar ansatz for the Dirac equation in the Kerr-Newman metric. Based on this, we derive an integral representation for smooth compactly supported functions which in turn we use to derive an integral representation for the propagator of solutions of the Cauchy problem with initial data in the above class of functions. As a by-product, we also obtain the propagator for the Dirac equation in the Minkowski space-time in oblate spheroidal coordinates. 
\end{abstract}


\maketitle

\section{Introduction}\label{sec:1}
One of the most spectacular predictions of general relativity are black holes which should form when a large mass is concentrated in a sufficiently small volume. The idea of a mass-concentration which is so dense that even light would be trapped goes back to Laplace in the 18th century. Shortly after Einstein developed general relativity, Karl Schwarzschild discovered in 1916 a mathematical solution to the equations of the theory that describes such an object. It was only much later, with the work of physicists like Oppenheimer, Volkoff and Snyder in the 1930's, that the scientific community began to think seriously about the possibility that such objects might actually exist in the Universe. It was shown that when a sufficiently massive star runs out of fuel, it is unable to support itself against its own gravitational attraction and it should collapse into a black hole. Starting with the 1960's and the 1970's, in the so-called Golden Era of black hole research, new interesting phenomena like the Hawking radiation and superradiance were discovered but for their rigorous mathematical description we have to wait until the 1990's and the beginning of the new century when the rigorous analysis of the propagation and of the scattering properties of classical and quantum fields on black hole space-times started to be developed.\\
Whenever we attempt to analyze the scattering properties of fields in the more general framework of the Kerr-Newman black hole geometry, we are faced with several difficulties which are not present in the picture of the Schwarzschild metric. First of all, the Kerr-Newman solution is only axially symmetric (cylindrical symmetry) since it possesses only two commuting Killing vector fields, namely the time coordinate vector field $\partial_{t}$ and the longitude coordinate vector field $\partial_{\varphi}$. This implies that there is no decomposition in spin-weighted spherical harmonics. Moreover, another difficulty is due to fact that the Kerr-Newman space-time is not stationary. In particular it is impossible to find a Killing vector field which is time-like everywhere outside the black hole. In fact $\partial_{t}$ becomes space-like in the ergo-sphere, a toroidal region around the horizon. This implies that for field equations describing particles of integer spin (wave equation, Klein-Gordon, Maxwell) there exists no positive definite conserved energy. For field equations describing particles with half integer spin (Weyl, Dirac) we can find a conserved $L_{2}$ norm with the usual interpretation of a conserved charge. Hence, the absence of stationarity in the Kerr-Newman metric is not really a difficulty for the scattering theory of classical Dirac fields.\\
For the reasons mentioned above there are only few analytical studies of the propagation of fields outside a Kerr-Newman black hole. Time-dependent scattering for the Klein-Gordon equation in the Kerr framework has been developed by H\"{a}fner (2003), while Daud$\acute{\mbox{e}}$ (2004) proved the existence and asymptotic completeness of wave operators, classical at the event horizon and Dollard-modified at infinity, for classical massive Dirac particles in the Kerr-Newman geometry.\\ 
Our present work is the first of two papers devoted to develop a time-dependent scattering theory for massive Dirac particles outside the event horizon of a non-extreme Kerr-Newman black hole. Before describing the contents of our work we should mention that the completeness of the Chandrasekhar ansatz has been proved for the first time in Finster et al. (2003). There, they derived an integral representation for the Dirac propagator in terms of the solutions of the radial and angular ODEs arising by means of the Chandrasekhar separation of variables which in turn they used to read out the completeness of the ansatz. In the present work we proceed in the opposite way. The main advantage is that the derivation of the propagator will follow quite immediately after we have proved the completeness of the Chandrasekhar ansatz. As a by-product, we also obtain an integral representation for the Dirac propagator in the Minkowski space-time in oblate spheroidal coordinates. Our integral representations together with the estimates for the asymptotic behaviour of the solutions of the radial problem will allow us in the second paper to give explicit analytical expressions for wave operators classical at the event horizon and Dollard-modified at infinity. Asymptotically away from the black hole it will turn out to be of advantage to consider first classical wave operators since their computation will lead us to an analytical expression for the time-dependent logarithmic phase shift needed to construct the Dollard-modified wave operators. After implementation of this phase shift in the free dynamics we will evaluate the Dollard-modified wave operators and obtain an integral representation for them. To our knowledge this will be the first analytical result since all previous works in this direction (see H\"{a}fner (2003) and Daud$\acute{\mbox{e}}$ (2004)) which are based on the Mourre theory (Mourre (1981)), treat only the problem of the existence of the above mentioned wave operators.\\
The present article is organised as follows: In Section 1 we motivate our interest in the derivation of an integral representation for the Dirac propagator in the Kerr-Newman metric. In Section~\ref{sec:2} we give a short review of the Dirac equation in the Kerr-Newman geometry. In Section~\ref{sec:3} we introduce the so-called Chandrasekhar ansatz. In Section~\ref{sec:4} we compute the scalar product with respect to which the Dirac Hamiltonian obtained from (\ref{uno}) is formally self-adjoint. In Section~\ref{sec:5} we bring the modified but equivalent Dirac equation (\ref{vdoppio}) into the form of a Schr\"{o}dinger equation with Hamiltonian $h$ and we show that $h$ defined on the set of the smooth functions with compact support is essentially self-adjoint. In Section~\ref{sec:6} the main results are Theorem~\ref{completo} where we prove the completeness of the Chandrasekhar ansatz and Theorem~\ref{Dirprop} which gives the integral representation for the Dirac propagator in a Kerr-Newman black hole manifold. Lemma~\ref{faraway} and ~\ref{eventhorizonasympt} in Section~\ref{sec:7} describe the asymptotic behaviour of the radial functions satisfying the system of first order ODEs (\ref{radial}) introduced in Section~\ref{sec:3}. Finally, in the last section we also derive an integral representation for the Dirac equation in the Minkowski space-time in oblate spheroidal coordinates.

\section{Preliminaries}
\subsection{The Dirac equation in the Newman-Penrose formalism}\label{sec:2}
According to Penrose and Rindler in terms of two-component spinors $(\phi^{A},\chi^{A^{'}})$ the Dirac equation coupled to a general gravitational field and a $4$-vector field $\textbf{V}$ is given in Planck units $\hbar=c=G=1$ by 
\begin{equation} 
(\nabla^{A}_{A^{'}}-ieV^{A}_{A^{'}})\phi_{A}=\frac{m_{e}}{\sqrt{2}}\chi_{A^{'}},\qquad   
(\nabla_{A}^{A^{'}}-ieV_{A}^{A^{'}})\chi_{A^{'}}=\frac{m_{e}}{\sqrt{2}}\phi_{A}  \label{unopage} 
\end{equation}
where $\nabla_{AA^{'}}$ is the symbol for covariant differentiation, $e$ is the charge or coupling constant of the Dirac particle to the vector field $\textbf{V}$ and $m_{e}$ is the particle mass. Notice that the factor $2^{-\frac{1}{2}}$ appearing in the above equations is due to the fact that the Pauli matrices as defined in the Newman-Penrose formalism differ from their usual definitions by the factor $\sqrt{2}$. The next step is to bring (\ref{unopage}) into a system of partial differential equations with respect to a coordinate basis. In the framework of Newman and Penrose this is achieved by choosing a null tetrad, i.e. a set of four vector fields $(\mathbf{l},\mathbf{n},\mathbf{m},\overline{\mathbf{m}})$ where $(\mathbf{l},\mathbf{n})$ is a pair of real vectors and $(\mathbf{m},\overline{\mathbf{m}})$ a pair of complex conjugate vectors. Moreover, they satisfy the orthogonality conditions $\mathbf{l}\cdot\mathbf{m}=\mathbf{l}\cdot\overline{\mathbf{m}}=\mathbf{n}\cdot\mathbf{m}=\mathbf{n}\cdot\overline{\mathbf{m}}=0$, the requirements that the vectors be null, i.e. $\mathbf{l}\cdot\mathbf{l}=\mathbf{n}\cdot\mathbf{n}=\mathbf{m}\cdot\mathbf{m}=\overline{\mathbf{m}}\cdot\overline{\mathbf{m}}=0$ and finally, the normalization conditions $\mathbf{l}\cdot\mathbf{n}=1$ and $\mathbf{m}\cdot\overline{\mathbf{m}}=-1$. The covariant derivative is decomposed into directional covariant derivatives along the frame vectors which we denote by
\begin{equation} \label{dirder1}
D=l^{a}\nabla_{a},\qquad\widetilde{\Delta}=n^{a}\nabla_{a},\qquad\delta=m^{a}\nabla_{a},\qquad\overline{\delta}=\overline{m}^{a}\nabla_{a}.
\end{equation}
The spin coefficients can be written in terms of the Ricci rotation-coefficients (e.g. Chandrasekhar (1992)) by defining the frame vectors and their dual $1$-forms as follows 
\begin{eqnarray*}
&&l^{a}=e_{(1)}^{\hspace{4mm} a},\qquad n^{a}=e_{(2)}^{\hspace{4mm} a},\qquad m^{a}=e_{(3)}^{\hspace{4mm} a},\qquad\overline{m}^{a}=e_{(4)}^{\hspace{4mm} a},\\
&&l_{a}=e^{(1)}_{\hspace{4mm} a},\qquad n_{a}=e^{(2)}_{\hspace{4mm} a},\qquad m_{a}=e^{(3)}_{\hspace{4mm} a},\qquad\overline{m}_{a}=e^{(4)}_{\hspace{4mm} a}
\end{eqnarray*}
where enclosure in parenthesis distinguishes the tetrad indices from the tensor indices. The Ricci rotation-coefficients are expressed by
\begin{eqnarray*}
&&\gamma_{(a)(b)(c)}=\frac{1}{2}\left[\lambda_{(a)(b)(c)}+\lambda_{(c)(a)(b)}-\lambda_{(b)(c)(a)}\right]\\
&&\lambda_{(a)(b)(c)}=e_{(b)i;j}\left[e_{(a)}^{\hspace{4mm} i}e_{(c)}^{\hspace{4mm} j}-e_{(a)}^{\hspace{4mm} j}e_{(c)}^{\hspace{4mm} i}\right],\qquad e_{(b)i}=g_{ik}e_{(b)}^{\hspace{4mm} k}
\end{eqnarray*}
where $g$ is the metric tensor. The spin coefficients can be written by means of the $\gamma_{(a)(b)(c)}$ as follows
\numparts
\begin{eqnarray} \label{s}
&&\kappa=\gamma_{(3)(1)(1)},\qquad \rho=\gamma_{(3)(1)(4)},\qquad \epsilon=\frac{1}{2}\left[\gamma_{(2)(1)(1)}+\gamma_{(3)(4)(1)}\right], \\
&&\sigma=\gamma_{(3)(1)(3)},\qquad \mu=\gamma_{(2)(4)(3)},\qquad \gamma=\frac{1}{2}\left[\gamma_{(2)(1)(2)}+\gamma_{(3)(4)(2)}\right], \\
&&\lambda=\gamma_{(2)(4)(4)},\qquad \tau=\gamma_{(3)(1)(2)},\qquad \alpha=\frac{1}{2}\left[\gamma_{(2)(1)(4)}+\gamma_{(3)(4)(4)}\right], \\
&&\nu=\gamma_{(2)(4)(2)},\qquad \pi=\gamma_{(2)(4)(1)},\qquad \beta=\frac{1}{2}\left[\gamma_{(2)(1)(3)}+\gamma_{(3)(4)(3)}\right]. 
\end{eqnarray}
\endnumparts 
Let us recall that to any tetrad $(\mathbf{l},\mathbf{n},\mathbf{m},\overline{\mathbf{m}})$ we can associate a unitary spin-frame $(o^{A},\iota^{A})$, defined uniquely up to an overall sign factor by the requirements that (for details see Penrose and Rindler)
\[
o^{A}\overline{o}^{A^{'}}=l^{a},\quad\iota^{A}\overline{\iota}^{A^{'}}=n^{a},\quad o^{A}\overline{\iota}^{A^{'}}=m^{a},\quad\iota^{A}\overline{o}^{A^{'}}=\overline{m}^{a},\quad o_{A}\iota^{A}=1.
\] 
Denoting by $\phi_{0}$ and $\phi_{1}$ the components of $\phi^{A}$ in $(o^{A},\iota^{A})$ and by $\chi_{0^{'}}$ and $\chi_{1^{'}}$ the components of $\chi^{A^{'}}$ in $(\overline{o}^{A^{'}},\overline{\iota}^{A^{'}})$
\[
\phi_{0}=\phi_{A}o^{A},\quad\phi_{1}=\phi_{A}\iota^{A},\quad\chi_{0^{'}}=\chi_{A^{'}}\overline{o}^{A^{'}},\quad\chi_{1^{'}}=\chi_{A^{'}}\overline{\iota}^{A^{'}},
\]
the Dirac equation (\ref{unopage}) takes the form
\numparts
\begin{eqnarray} \label{dirac1}
&&(D+\overline{\epsilon}-\overline{\rho}-iel^{a}V_{a})\chi_{1^{'}}-(\delta+\overline{\pi}-\overline{\alpha}-iem^{a}V_{a})\chi_{0^{'}}=-\frac{m_{e}}{\sqrt{2}}\phi_{0},\\
&&(\widetilde{\Delta}+\overline{\mu}-\overline{\gamma}-ien^{a}V_{a})\chi_{0^{'}}-(\overline{\delta}+\overline{\beta}-\overline{\tau}-ie\overline{m}^{a}V_{a})\chi_{1^{'}}=\frac{m_{e}}{\sqrt{2}}\phi_{1},\\
&&(\widetilde{\Delta}+\mu-\gamma-ien^{a}V_{a})\phi_{0}-(\delta+\beta-\tau-iem^{a}V_{a})\phi_{1}=\frac{m_{e}}{\sqrt{2}}\chi_{1^{'}},\\
&&(D+\epsilon-\rho-iel^{a}V_{a})\phi_{1}-(\overline{\delta}+\pi-\alpha-ie\overline{m}^{a}V_{a})\phi_{0}=-\frac{m_{e}}{\sqrt{2}}\chi_{0^{'}}.
\end{eqnarray}
\endnumparts
Let us introduce the following symbols
\begin{eqnarray*}
\hspace{-3mm}&&\alpha_{+}=-i\sqrt{2}(D+\overline{\epsilon}-\overline{\rho}-iel^{a}V_{a}),\quad\hspace{2mm}\beta_{+}=i\sqrt{2}(\delta+\overline{\pi}-\overline{\alpha}-iem^{a}V_{a}),\\
\hspace{-3mm}&&\beta_{-}=i\sqrt{2}(\overline{\delta}+\overline{\beta}-\overline{\tau}-ie\overline{m}^{a}V_{a}),\quad\hspace{3.5mm}\alpha_{-}=-i\sqrt{2}(\widetilde{\Delta}+\overline{\mu}-\overline{\gamma}-ien^{a}V_{a}),\\
\hspace{1mm}&&\widetilde{\alpha}_{-}=-i\sqrt{2}(\widetilde{\Delta}+\mu-\gamma-ien^{a}V_{a}),\quad\hspace{0.5mm}\widetilde{\beta}_{+}=i\sqrt{2}(\delta+\beta-\tau-iem^{a}V_{a}),\\
\hspace{-3mm}&&\widetilde{\beta}_{-}=i\sqrt{2}(\overline{\delta}+\pi-\alpha-ie\overline{m}^{a}V_{a}),\quad\hspace{3.5mm}\widetilde{\alpha}_{+}=-i\sqrt{2}(D+\epsilon-\rho-iel^{a}V_{a})
\end{eqnarray*}
and let us define the spinor $\Psi=(F_{1},F_{2},G_{1},G_{2})^{T}$ with components
\[
F_{1}=-\phi_{0},\quad F_{2}=\phi_{1},\quad G_{1}=i\chi_{1^{'}},\quad G_{2}=i\chi_{0^{'}}.
\]
Then the system of equations (2.4{\it{a}})-(2.4{\it{d}}) is simply
\begin{equation} \label{uno}
\mathcal{O}_{D}\Psi=\left(\begin{array}{cccc} 
-m_{e}&0&\alpha_{+}&\beta_{+}\\
0&-m_{e}&\beta_{-}&\alpha_{-}\\
\widetilde{\alpha}_{-}&-\widetilde{\beta}_{+}&-m_{e}&0\\
-\widetilde{\beta}_{-}&\widetilde{\alpha}_{+}&0&-m_{e}
\end{array}\right)\Psi=0.
\end{equation}
In what follows, we consider the Dirac equation in the Kerr-Newman geometry, i.e. in the presence of a rotating charged black hole. In Boyer-Lindquist coordinates $(t,r,\vartheta,\varphi)$ with $r>0$, $0\leq\vartheta\leq\pi$, $0\leq\varphi<2\pi$ the Kerr-Newman metric is given by (e.g. Wald)
\[
ds^{2}=\frac{\Delta-a^{2}\sin^{2}{\vartheta}}{\Sigma}dt^{2}+\frac{2a\sin^{2}{\vartheta}(r^2+a^2-\Delta)}{\Sigma}dtd\varphi
\]
\begin{equation}\label{KN}
\hspace{5cm}-\frac{\Sigma}{\Delta}dr^2-\Sigma d\vartheta^{2}-(r^2+a^2)^{2}\sin^{2}{\vartheta}\frac{\widetilde{\Sigma}}{\Sigma}d\varphi^{2}
\end{equation}
with
\[
\Sigma:=\Sigma(r,\theta)=r^2+a^2\cos^{2}\theta, \qquad \Delta:=\Delta(r)=r^2-2Mr+a^2+Q^2
\]
and
\[
\widetilde{\Sigma}:=\widetilde{\Sigma}(r,\vartheta)=1-a^2\gamma^{2}(r)\sin^{2}{\vartheta},\qquad \gamma(r):=\frac{\sqrt{\Delta}}{r^2+a^2}
\]
where $M$, $a$ and $Q$ are the mass, the angular momentum per unit mass and the charge of the black hole, respectively. Here we will always consider $a\neq 0$. In the non-extreme case $M^2>a^2+Q^2$ the function $\Delta$ has two distinct zeros, namely,
\[ 
r_{0}=M-\sqrt{M^2-a^2-Q^2}, \qquad r_{1}=M+\sqrt{M^2-a^2-Q^2},
\]
the first one corresponding to the Cauchy horizon and the second to the event horizon. Moreover, $\Delta>0$ for $r>r_{1}$. In the extreme case $M^2=a^2+Q^2$ the Cauchy horizon and the event horizon coincide since $\Delta$ has a double root at $r_{1}^{*}=M$.
\begin{lemma} \label{lemma0}
Let $M^{2}\geq a^{2}+Q^{2}$. For every $r>r_{1}>0$ and $\vartheta\in[0,\pi]$ it results $\widetilde{\Sigma}>0$.
\end{lemma}
\begin{proof}
Let us write $\widetilde{\Sigma}$ as follows
\[
\widetilde{\Sigma}=\frac{\widetilde{p}(r,\vartheta)}{(r^{2}+a^{2})^{2}},\quad \widetilde{p}(r,\vartheta)=(r^{2}+a^{2})^{2}-a^{2}\Delta\sin^{2}{\vartheta}
\]
A short computation gives
\[
\widetilde{p}(r,\vartheta)=r^{4}+a^{2}(1+\cos^{2}{\vartheta})r^{2}+2Ma^{2}r\sin^{2}{\vartheta}+a^{2}(a^{2}\cos^{2}{\vartheta}-Q^{2}\sin^{2}{\vartheta})
\]
where we used the definition of $\Delta$. Since $r>r_{1}>M$, we have
\begin{eqnarray*}
\hspace{-5mm}\widetilde{p}(r,\vartheta)&>&M^{4}+a^{2}M^{2}(1+\cos^{2}{\vartheta})+2M^{2}a^{2}\sin^{2}{\vartheta}+a^{2}(a^{2}\cos^{2}{\vartheta}-Q^{2}\sin^{2}{\vartheta}),\\
    &=&M^{4}+2a^{2}M^{2}+a^{4}\cos^{2}{\vartheta}+a^{2}(M^{2}-Q^{2})\sin^{2}{\vartheta}.
\end{eqnarray*}
Recall that in the non-extreme case $M^{2}>Q^{2}$, while in the extreme case $M^{2}-Q^{2}=a^{2}$. Hence, the last line in the above expression is positive for every $\vartheta\in[0,\pi]$.\hspace{5mm}\opensquare
\end{proof}
Finally, notice that by setting $M=Q=0$ in (\ref{KN}) the Kerr-Newman metric goes over to the Minkowski metric in oblate spheroidal coordinates (OSC), namely
\begin{equation} \label{Mink}
ds^{2}=dt^{2}-\frac{\Sigma}{\widehat{\Delta}}dr^{2}-\Sigma d\vartheta^{2}-\widehat{\Delta}\sin^{2}\vartheta d\varphi^{2},\quad \widehat{\Delta}=r^2+a^2.
\end{equation}
In fact, by means of the coordinate transformation
\begin{equation} \label{OSCco}
x=\sqrt{r^2+a^2}\sin{\vartheta}\cos{\varphi},\quad y=\sqrt{r^2+a^2}\sin{\vartheta}\sin{\varphi},\quad z=r\cos{\vartheta}
\end{equation}
(\ref{Mink}) can be reduced to the Minkowski metric in cartesian coordinates. In the case $M=Q=0$ the function $\widehat{\Delta}$ has no real zeros. Thus there are no coordinate singularities. However the metric has a curvature singularity at $\Sigma=0$, i.e. for $r=0$ and $\vartheta=\frac{\pi}{2}$. The surfaces of constant $r$ are confocal ellipsoids which degenerate at $r=0$ to the disc $z=0$ and $x^{2}+y^{2}\leq a^{2}$. The surfaces of constant $\vartheta$ are represented by hyperboloids of one sheet, confocal to the above mentioned ellipsoids and $\vartheta=\frac{\pi}{2}$ corresponds to the boundary of the disc at $x^{2}+y^{2}=a^{2}$. As a consequence the curvature singularity occurs on the boundary of the disc, i.e. on the circle $x^{2}+y^{2}=a^{2}$ and $z=0$. Moreover, there is no reason to restrict $r$ to be positive since the space-time can be analytically continued through the disc to another flat region with $r<0$. For further details we refer to Flammer (1957) and to the original work of Boyer and Lindquist (1967).\\
The Dirac equation in the Kerr-Newman geometry was computed and separated by Page (1976) with the help of the Kinnersley tetrad (see Kinnersley (1969)). We recall that such a tetrad is naturally inherited from the Petrov type D structure of the Kerr-Newman metric (e.g. O'Neill (1995)). In view of the separation of the Dirac equation we choose to work with the Carter tetrad (Carter (1987)) 
\numparts
\begin{eqnarray} \label{t1}
l^{a}\nabla_{a}&=&\frac{1}{\sqrt{2\Delta\Sigma}}\left(r^{2}\frac{\partial}{\partial\widetilde{t}}-\Delta\frac{\partial}{\partial r}+\frac{\partial}{\partial\widetilde{\varphi}}\right)\\
n^{a}\nabla_{a}&=&\frac{1}{\sqrt{2\Delta\Sigma}}\left(r^{2}\frac{\partial}{\partial\widetilde{t}}+\Delta\frac{\partial}{\partial r}+\frac{\partial}{\partial\widetilde{\varphi}}\right),\\ 
m^{a}\nabla_{a}&=&\frac{1}{\sqrt{2(a^{2}-q^{2})\Sigma}}\left(iq^{2}\frac{\partial}{\partial\widetilde{t}}-(a^{2}-q^{2})\frac{\partial}{\partial q}-i\frac{\partial}{\partial\widetilde{\varphi}}\right)
\end{eqnarray}
\endnumparts
where the dependency of the new variables $\widetilde{t}$, $q$ and $\widetilde{\varphi}$ on the Boyer-Lindquist variables is given by the relations
\[
\widetilde{t}=t-a\varphi,\quad q=a\cos{\vartheta},\qquad \widetilde{\varphi}=\frac{\varphi}{a}.
\]  
In the above defined variables the electromagnetic potential is simply given by
\[
V_{a}=\frac{Qr}{\Sigma}(1,0,0,q^{2}),\quad \Sigma=r^{2}+q^{2}
\]
which is stationary and axially symmetric. Using the above tetrad and (\ref{s})-(2.3{\it{d}}) the spin coefficients are computed to be 
\begin{eqnarray*}
\hspace{-1.5cm}&&\kappa=\sigma=\lambda=\nu=0,\quad\rho=\mu,\quad \gamma=\epsilon,\quad\pi=-\tau,\quad\beta=-\alpha, \\
&&\mu=-\sqrt{\frac{\Delta}{2\Sigma}}\widetilde{\rho},\quad\epsilon=\frac{-\Delta\widetilde{\rho}-(r-M)}{2\sqrt{2\Delta\Sigma}},\quad\tau=-i\widetilde{\rho}\sqrt{\frac{a^{2}-q^{2}}{2\Sigma}},\\
&&\alpha=\frac{i(a^{2}-q^{2})r-(r^2+a^2)q}{2\sqrt{2(a^{2}-q^{2})\Sigma}}
\end{eqnarray*}
with $\widetilde{\rho}=-(r-iq)^{-1}$. Notice that the vanishing of $\kappa$ and $\sigma$ implies that the congruence formed by the vector $\textbf{l}$ is geodesic and shear-free, respectively. As a consequence the Riemann tensor is of type II according to the Goldberg-Sachs theorem (see Goldberg and Sachs (1962)).\\
We are now ready to give a more explicit form to the Dirac equation (\ref{uno}). Indeed, by introducing the following operators
\begin{eqnarray}
&&\mathcal{D}_{\pm}=\frac{\partial}{\partial r}\mp\frac{1}{\Delta}\left[(r^2+a^2)\frac{\partial}{\partial t}+a\frac{\partial}{\partial\varphi}-ieQr\right],\label{DPM}\\
&&\mathcal{L}_{\pm}=\frac{\partial}{\partial \vartheta}+\frac{1}{2}\cot{\vartheta}\mp i\left(a\sin{\vartheta}\frac{\partial}{\partial t}+\csc{\vartheta}\frac{\partial}{\partial\varphi}\right)\label{LPM},
\end{eqnarray}
the entries of the matrix in (\ref{uno}) are computed to be
\numparts
\begin{eqnarray} \label{alphabeta}
&&\alpha_{\pm}=\pm i\sqrt{\frac{\Delta}{\Sigma}}\left(\mathcal{D}_{\pm}+f(r,\vartheta)\right),\quad\beta_{\pm}=\frac{i}{\sqrt{\Sigma}}\left(\mathcal{L}_{\pm}+g(r,\vartheta)\right),\\
&&\widetilde{\alpha}_{\pm}=\pm i\sqrt{\frac{\Delta}{\Sigma}}\left(\mathcal{D}_{\pm}+\overline{f}(r,\vartheta)\right),\quad\widetilde{\beta}_{\pm}=\frac{i}{\sqrt{\Sigma}}\left(\mathcal{L}_{\pm}+\overline{g}(r,\vartheta)\right)
\end{eqnarray}
\endnumparts
with
\begin{equation}\label{fg}
f(r,\vartheta)=\frac{1}{2}\left(\frac{r-M}{\Delta}+\frac{1}{r+ia\cos{\vartheta}}\right),\quad g(r,\vartheta)=-\frac{ia\sin{\vartheta}}{2(r+ia\cos{\vartheta})}.
\end{equation}

\subsection{The Chandrasekhar separation ansatz}\label{sec:3}
Carter and McLenaghan (1979) found that in the Kerr geometry and in the flat-space limit there exists a generalized total angular momentum operator of the form
\begin{equation}\label{L}
J=i\gamma_{5}\gamma^{\mu}\left(f_{\mu}^{\nu}D_{\nu}-\frac{1}{6}\gamma^{\nu}\gamma^{\rho}f_{\mu\nu;\rho}\right),\quad D_{\nu}=\nabla_{\nu}-ieA_{\nu},
\end{equation}
commuting with the Dirac operator under the condition that $f_{\mu}^{\nu}$ is an antisymmetric tensor satisfying the Penrose-Floyd equation $f_{\mu(\nu;\rho)}=0$ (see Penrose and Floyd (1973)). In particular Carter and McLenaghan  also showed that (\ref{L}) can be interpreted in the Minkowski space as the square root of the ordinary total squared angular momentum Casimir operator of the rotation group. In this section we recall how to construct from the Chandrasekhar ansatz (see Chandrasekhar (1976)) for the Dirac equation in the Kerr-Newman metric the symmetry operator $J$ for the formal Dirac operator $\mathcal{O}_{D}$ defined in (\ref{uno}) and we compute it explicitly. As pointed out in Carter and McLenaghan the existence of such an operator underlaying the Chandrasekhar separability of the Dirac equation is due to the presence of an appropriate Killing spinor field on the space-time under consideration. For the concept of Killing spinor we refer to the original work of Penrose and Walker (1970).\\
We begin by replacing the Dirac equation (\ref{uno}) by a modified but equivalent equation
\begin{equation} \label{vdoppio}
\mathcal{W}\widehat{\psi}=(\Gamma S^{-1}\mathcal{O}_{D}S)\widehat{\psi}=0
\end{equation}
where $\widehat{\psi}=S^{-1}\Psi=(\widehat{F}_{1},\widehat{F}_{2},\widehat{G}_{1},\widehat{G}_{2})^{T}$ and $\Gamma$ and $S$ are non singular $4\times 4$ matrices, whose elements may depend on the variables $r$ and $\vartheta$. The next lemma shows how to find non singular matrices $S$ and $\Gamma$ such that the formal operator $\mathcal{W}$ decomposes into the sum of two formal operators, the first one containing only derivatives respect to the variables $t$, $r$ and $\varphi$ and the second one involving only derivatives respect to $t$, $\vartheta$ and $\varphi$. 
\begin{lemma} \label{lemma1}
Let $r$ be positive with $r>r_{1}$. Moreover, let $S$ and $\Gamma$ be non singular matrices defined as follows 
\begin{equation}\label{trafS}
\hspace{-1.5cm}S=\Delta^{-\frac{1}{4}}~\mbox{\upshape{diag}}\left(\frac{1}{\sqrt{r-ia\cos{\vartheta}}},\frac{1}{\sqrt{r-ia\cos{\vartheta}}}, \frac{1}{\sqrt{r+ia\cos{\vartheta}}},\frac{1}{\sqrt{r+ia\cos{\vartheta}}}\right)
\end{equation}
with $\mbox{\upshape{det}}(S)=(\Sigma\Delta)^{-1}$ and 
\[
\hspace{-1.5cm}\Gamma=-i~\mbox{\upshape{diag}}(r+ia\cos{\vartheta},-(r+ia\cos{\vartheta}),-(r-ia\cos{\vartheta}),r-ia\cos{\vartheta})
\]
with $\mbox{\upshape{det}}(\Gamma)=\Sigma^2$. Then
\begin{equation} \label{unoo}
\mathcal{W}=\mathcal{W}_{(t,r,\varphi)}+\mathcal{W}_{(t,\vartheta,\varphi)}
\end{equation}
where
\begin{equation} \label{due}
\mathcal{W}_{(t,r,\varphi)}=\left( \begin{array}{cccc}
                            im_{e}r&0&\sqrt{\Delta}\mathcal{D}_{+}&0\\
                            0&-im_{e}r&0&\sqrt{\Delta}\mathcal{D}_{-}\\
                            \sqrt{\Delta}\mathcal{D}_{-}&0&-im_{e}r&0\\
                             0&\sqrt{\Delta}\mathcal{D}_{+}&0&im_{e}r
                            \end{array} \right),
\end{equation}
\begin{equation} \label{3}
\mathcal{W}_{(t,\vartheta,\varphi)}=\left( \begin{array}{cccc}
                            -am_{e}\cos{\vartheta}&0&0&\mathcal{L}_{+}\\
                            0&am_{e}\cos{\vartheta}&-\mathcal{L}_{-}&0\\
                            0&\mathcal{L}_{+}&-am_{e}\cos{\vartheta}&0\\
                            -\mathcal{L}_{-}&0&0&am_{e}\cos{\vartheta}
                            \end{array} \right)
\end{equation}
with $\mathcal{D}_{\pm}$ and $\mathcal{L}_{\pm}$ defined by (\ref{DPM}) and (\ref{LPM}) in Section~\ref{sec:2}.
\end{lemma}
\begin{proof}
Let us introduce an invertible matrix
\[
S:=\mbox{\upshape{diag}}(h(r,\vartheta),\Lambda(r,\vartheta),\sigma(r,\vartheta),\gamma(r,\vartheta))
\]
with functions $h$, $\Lambda$, $\sigma$, $\gamma$ at least $\mathcal{C}^{1}((r_{1},+\infty)\times[0,\pi])$. Then the equation $\mathcal{O}_{D}S\widehat{\psi}=0$ gives rise to the system
\numparts
\begin{eqnarray} 
&&-m_{e}h\widehat{F}_{1}+\alpha_{+}(\sigma\widehat{G}_{1})+\beta_{+}(\gamma\widehat{G}_{2})=0,\label{A}\\
&&-m_{e}\Lambda\widehat{F}_{2}+\beta_{-}(\sigma\widehat{G}_{1})+\alpha_{-}(\gamma\widehat{G}_{2})=0,\label{AA1}\\
&&-m_{e}\sigma\widehat{G}_{1}+\widetilde{\alpha}_{-}(h\widehat{F}_{1})-\widetilde{\beta}_{+}(\Lambda\widehat{F}_{2})=0 ,\\
&&-m_{e}\gamma\widehat{G}_{2}-\widetilde{\beta}_{-}(h\widehat{F}_{1})+\widetilde{\alpha}_{+}(\Lambda\widehat{F}_{2})=0.
\end{eqnarray}
\endnumparts
If we substitute (\ref{alphabeta})-(2.12{\it{b}}), we obtain
\numparts
\begin{eqnarray}\label{A1}
&&\alpha_{+}(\sigma\widehat{G}_{1})=i\sqrt{\frac{\Delta}{\Sigma}}\left(\sigma\mathcal{D}_{+}+\frac{\partial\sigma}{\partial r}+f\sigma\right)\widehat{G}_{1},\\
&&\beta_{+}(\gamma\widehat{G}_{2})=\frac{i}{\sqrt{\Sigma}}\left(\gamma\mathcal{L}_{+}+\frac{\partial\gamma}{\partial \vartheta}+g\gamma\right)\widehat{G}_{2}.
\end{eqnarray}
\endnumparts
Let $\sigma$ and $\gamma$ be
\[
\sigma(r,\theta)=\frac{\widetilde{\sigma}(\vartheta)}{\Delta^{\frac{1}{4}}\sqrt{r+ia\cos{\vartheta}}},\quad\gamma(r,\vartheta)=\frac{\widetilde{\gamma}(r)}{\sqrt{r+ia\cos{\vartheta}}}
\]
with the constraints $\widetilde{\sigma}(\vartheta)\neq 0$ for every $\vartheta\in[0,\pi]$ and $\widetilde{\gamma}(r)\neq 0$ for every $r\in(r_{1},+\infty)$. Then (\ref{A1}\,\textit{-b}) simplify to
\[
\alpha_{+}(\sigma\widehat{G}_{1})=i\sqrt{\frac{\Delta}{\Sigma}}\sigma\mathcal{D}_{+}\widehat{G}_{1},\quad\beta_{+}(\gamma\widehat{G}_{2})=\frac{i}{\sqrt{\Sigma}}\gamma\mathcal{L}_{+}\widehat{G}_{2}.
\]
The substitution of (\ref{alphabeta})-(2.12{\it{b}}) in (\ref{AA1}) gives
\numparts
\begin{eqnarray}\label{A3}
&&\alpha_{-}(\gamma\widehat{G}_{2})=-i\sqrt{\frac{\Delta}{\Sigma}}\left(\gamma\mathcal{D}_{-}+\frac{\partial\gamma}{\partial r}+f\gamma\right)\widehat{G}_{2},\\
&&\beta_{-}(\sigma\widehat{G}_{1})=\frac{i}{\sqrt{\Sigma}}\left(\sigma\mathcal{L}_{-}+\frac{\partial\sigma}{\partial \vartheta}+g\sigma\right)\widehat{G}_{1}.
\end{eqnarray}
\endnumparts
It is easy to verify that (\ref{A3})-(2.22{\it{b}}) reduce to
\[
\alpha_{-}(\gamma\widehat{G}_{2})=-i\sqrt{\frac{\Delta}{\Sigma}}~\gamma~\mathcal{D}_{-}\widehat{G}_{2},\quad\beta_{-}(\sigma\widehat{G}_{1})=\frac{i}{\sqrt{\Sigma}}~\sigma\mathcal{L}_{-}\widehat{G}_{1}
\]
by choosing $\widetilde{\gamma}(r)=c_{4}\Delta^{-\frac{1}{4}}$ with $c_{4}\in\mathbb{C}\backslash\{0\}$ and $\widetilde{\sigma}(\vartheta)=c_{3}\in\mathbb{C}\backslash\{0\}$. Analogously we find that
\[
h(r,\vartheta)=\frac{c_{1}}{\Delta^{\frac{1}{4}}\sqrt{r-ia\cos{\vartheta}}},\quad\Lambda(r,\vartheta)=\frac{c_{2}}{\Delta^{\frac{1}{4}}\sqrt{r+ia\cos{\vartheta}}}
\]
for some $c_{1},$ $c_{2}\in\mathbb{C}\backslash\{0\}$. Let us define a non singular $4\times 4$ matrix 
\[
\Gamma=-i~\mbox{\upshape{diag}}(r+ia\cos{\vartheta},-(r+ia\cos{\vartheta}),-(r-ia\cos{\vartheta}),r-ia\cos{\vartheta}).
\]
After a tedious computation we find that
\[ 
\hspace{-1.2cm}\mathcal{W}=\Gamma S^{-1}\mathcal{O}S=
\left( \begin{array}{cccc}
\scriptstyle im_{e}r-am_{e}\cos{\vartheta}&0&\frac{c_{3}}{c_{1}}\scriptstyle\sqrt{\Delta}\mathcal{D}_{+}&\frac{c_{4}}{c_{1}}\scriptstyle\mathcal{L}_{+}\\
\scriptstyle 0&\scriptstyle -im_{e}r+am_{e}\cos{\vartheta}&-\frac{c_{3}}{c_{2}}\scriptstyle \mathcal{L}_{-}&\frac{c_{4}}{c_{2}}\scriptstyle \sqrt{\Delta}\mathcal{D}_{-}\\
\frac{c_{1}}{c_{3}}\scriptstyle \sqrt{\Delta}\mathcal{D}_{-}&\frac{c_{2}}{c_{3}}\scriptstyle \mathcal{L}_{+}&\scriptstyle -im_{e}r-am_{e}\cos{\vartheta}&\scriptstyle 0\\
-\frac{c_{1}}{c_{4}}\scriptstyle \mathcal{L}_{-}&\frac{c_{2}}{c_{4}}\scriptstyle \sqrt{\Delta}\mathcal{D}_{+}&\scriptstyle 0&\scriptstyle im_{e}r+am_{e}\cos{\vartheta}
\end{array} \right).
\]
If we require that $c_{1}=c_{2}=c_{3}=c_{4}=:c$, we finally have
\[ 
\hspace{0.9cm}\mathcal{W}=\left( \begin{array}{cccc}
\scriptstyle im_{e}r-am_{e}\cos{\vartheta}&\scriptstyle 0&\scriptstyle \sqrt{\Delta}\mathcal{D}_{+}&\scriptstyle \mathcal{L}_{+}\\
\scriptstyle 0&\scriptstyle -im_{e}r+am_{e}\cos{\vartheta}&-\scriptstyle \mathcal{L}_{-}&\scriptstyle \sqrt{\Delta}\mathcal{D}_{-}\\
\scriptstyle \sqrt{\Delta}\mathcal{D}_{-}&\scriptstyle \mathcal{L}_{+}&\scriptstyle -im_{e}r-am_{e}\cos{\vartheta}&\scriptstyle 0\\
-\scriptstyle \mathcal{L}_{-}&\scriptstyle \sqrt{\Delta}\mathcal{D}_{+}&\scriptstyle 0&\scriptstyle im_{e}r+am_{e}\cos{\vartheta}
                           \end{array} \right).
\]
Without loss of generality we can set $c=1$ and this completes the proof.\hspace{5mm}\opensquare
\end{proof}
\begin{remark}
It should be noted that the regular and time-independent transformation $S$ (eq. (2.1) in Finster \etal. (2003)) is in contrast with our (\ref{trafS}).
\end{remark}
Notice that Lemma~\ref{lemma1} is still valid also in the extreme case where $\Delta=(r-M)^2$. Since by setting $M=Q=0$ (see discussion in Section~\ref{sec:2}) the Kerr-Newman metric becomes the Minkowski metric in oblate spheroidal coordinates, we obtain from Lemma~\ref{lemma1} an analogous result for the Dirac equation in OSC which we state in the following corollary.
 
\begin{corollary} \label{corollary1}
Let $\mathcal{R}$ denote the circle $\mathcal{R}=\{(x,y,z)\in\mathbb{R}^{3}|x^{2}+y^{2}=a^{2},\hspace{2mm}z=0\}$. Moreover, in $\mathbb{R}^{3}\backslash\mathcal{R}$ let us define the non singular matrices 
\[
\hspace{-4mm}\widetilde{S}=\widehat{\Delta}^{-\frac{1}{4}}~\mbox{\upshape{diag}}\left(\frac{1}{\sqrt{r-ia\cos{\vartheta}}},\frac{1}{\sqrt{r-ia\cos{\vartheta}}}, \frac{1}{\sqrt{r+ia\cos{\vartheta}}},\frac{1}{\sqrt{r+ia\cos{\vartheta}}}\right), 
\]
with $\widehat{\Delta}=r^{2}+a^{2}$ and $\Gamma$ as in Lemma~\ref{lemma1}. Then
\begin{eqnarray}
&&\mathcal{W}^{(OSC)}=\mathcal{W}_{(t,r,\varphi)}^{(OSC)}+\mathcal{W}_{(t,\vartheta,\varphi)},\label{unostern}\\
&&\mathcal{W}^{(OSC)}_{(t,r,\varphi)}=\left( \begin{array}{cccc}
                            im_{e}r&0&\sqrt{\widehat{\Delta}}\widetilde{\mathcal{D}}_{+}&0\\
                            0&-im_{e}r&0&\sqrt{\widehat{\Delta}}\widetilde{\mathcal{D}}_{-}\\
                            \sqrt{\widehat{\Delta}}\widetilde{\mathcal{D}}_{-}&0&-im_{e}r&0\\
                             0&\sqrt{\widehat{\Delta}}\widetilde{\mathcal{D}}_{+}&0&im_{e}r
                            \end{array} \right)
\end{eqnarray}
where $\mathcal{W}_{(t,\vartheta,\varphi)}$ is given by (\ref{3}) and
\[
\widetilde{\mathcal{D}}_{\pm}:=\frac{\partial}{\partial r}\mp\left(\frac{\partial}{\partial t}+\frac{a}{\widehat{\Delta}}\frac{\partial}{\partial\varphi}\right) .
\]
\end{corollary}
It is interesting to observe that the angular operator $\mathcal{W}_{(t,\vartheta,\varphi)}$ remains unchanged for the Dirac equation in the Kerr-Newman metric and in the Minkowski space-time in oblate spheroidal coordinates. We compute now the commutation relations which we will need later to verify that in the Kerr-Newman metric the operator $J$ given by (\ref{L}) is a symmetry operator for $\mathcal{O}_{D}$ defined in (\ref{uno}). It is straightforward to check that 
\begin{equation} \label{4}
\left[\mathcal{W}_{(t,r,\varphi)},\mathcal{W}_{(t,\vartheta,\varphi)}\right]=0,\quad\left[\mathcal{W}_{(t,r,\varphi)},\mathcal{W}\right]=\left[\mathcal{W}_{(t,\vartheta,\varphi)},\mathcal{W}\right]=0.
\end{equation} 
Furthermore, the matrix $\Gamma$ introduced in Lemma~\ref{lemma1} splits into the sum
\begin{equation} \label{cinque}
\Gamma=\Gamma_{(r)}+\Gamma_{(\vartheta)}
\end{equation}
with $\Gamma_{(r)}$ and $\Gamma_{(\vartheta)}$ satisfying the commutator relations
\begin{equation} \label{sei}
\left[\Gamma_{(r)},\Gamma_{(\vartheta)}\right]=0,\quad\left[\Gamma_{(r)},\mathcal{W}_{(t,\vartheta,\varphi)}\right]=\left[\Gamma_{(\vartheta)},\mathcal{W}_{(t,r,\varphi)}\right]=0.
\end{equation}
In what follows we restrict our attention to stationary solutions of the Dirac equation. In particular since the Kerr-Newman metric is axially symmetric, it is natural to make the following ansatz for the spinors $\widehat{\psi}$ entering in (\ref{vdoppio}) 
\begin{equation} \label{psi}
\widehat{\psi}(t,r,\vartheta,\varphi)=e^{i\omega t}e^{i\left(k+\frac{1}{2}\right)\varphi}\widetilde{\psi}(r,\vartheta)
\end{equation}
where $\omega$ and $k\in\mathbb{Z}$ denote the energy and the azimuthal quantum number of the particle, respectively. Let us substitute (\ref{psi}) into (\ref{vdoppio}). Then $\widetilde{\psi}(r,\vartheta)$ satisfies the equation
\begin{equation} \label{mod}
\left(\mathcal{W}_{(r)}+\mathcal{W}_{(\vartheta)}\right)\widetilde{\psi}=0
\end{equation}
where
\numparts
\begin{eqnarray} \label{mod1}
&&\mathcal{W}_{(r)}=\left( \begin{array}{cccc}
                            \scriptstyle im_{e}r&\scriptstyle 0&\scriptstyle \sqrt{\Delta}\widehat{\mathcal{D}}_{+}&\scriptstyle 0\\
                            \scriptstyle0&\scriptstyle-im_{e}r&\scriptstyle0&\scriptstyle\sqrt{\Delta}\widehat{\mathcal{D}}_{-}\\
                            \scriptstyle \sqrt{\Delta}\widehat{\mathcal{D}}_{-}&\scriptstyle 0&\scriptstyle -im_{e}r&\scriptstyle 0\\
                            \scriptstyle 0&\scriptstyle \sqrt{\Delta}\widehat{\mathcal{D}}_{+}&\scriptstyle 0&\scriptstyle im_{e}r
                            \end{array} \right),\\
&&\mathcal{W}_{(\vartheta)}=\left( \begin{array}{cccc}
                            \scriptstyle-am_{e}\cos{\vartheta}&\scriptstyle0&\scriptstyle0&\scriptstyle\widehat{\mathcal{L}}_{+}\\
                            \scriptstyle0&\scriptstyle am_{e}\cos{\vartheta}&\scriptstyle-\widehat{\mathcal{L}}_{-}&\scriptstyle0\\
                            \scriptstyle0&\scriptstyle\widehat{\mathcal{L}}_{+}&\scriptstyle-am_{e}\cos{\vartheta}&\scriptstyle0\\
                            \scriptstyle-\widehat{\mathcal{L}}_{-}&\scriptstyle0&\scriptstyle0&\scriptstyle am_{e}\cos{\vartheta}
                            \end{array} \right)
\end{eqnarray}
\endnumparts
with
\numparts
\begin{eqnarray}\label{ramba}
\hspace{-0.8cm}&&\widehat{\mathcal{D}}_{\pm}=\frac{\partial}{\partial r}\mp i\frac{K(r)}{\Delta},\quad \hspace{1.5cm}K(r)=\omega(r^2+a^2)-eQr+\left(k+\frac{1}{2}\right)a,\\
\hspace{-0.8cm}&&\widehat{\mathcal{L}}_{\pm}=\frac{\partial}{\partial \vartheta}+\frac{1}{2}\cot{\vartheta}\pm Q(\vartheta),\quad Q(\vartheta)=a\omega\sin{\vartheta}+\left(k+\frac{1}{2}\right)\csc{\vartheta}.
\end{eqnarray}
\endnumparts
If we set $\widetilde{\psi}(r,\vartheta)=(f_{1}(r,\vartheta),f_{2}(r,\vartheta),g_{1}(r,\vartheta),g_{2}(r,\vartheta))^{T}$, equation (\ref{mod}) gives rise to the following system of first order linear partial differential equations
\begin{eqnarray*}
\sqrt{\Delta}\widehat{\mathcal{D}}_{+}g_{1}+im_{e}rf_{1}+(\widehat{\mathcal{L}}_{+}g_{2}-am_{e}\cos{\vartheta}f_{1})&=&0,\\
\sqrt{\Delta}\widehat{\mathcal{D}}_{-}g_{2}-im_{e}rf_{2}-(\widehat{\mathcal{L}}_{-}g_{1}-am_{e}\cos{\vartheta}f_{2})&=&0,\\
\sqrt{\Delta}\widehat{\mathcal{D}}_{-}f_{1}-im_{e}rg_{1}+(\widehat{\mathcal{L}}_{+}f_{2}-am_{e}\cos{\vartheta}g_{1})&=&0,\\
\sqrt{\Delta}\widehat{\mathcal{D}}_{+}f_{2}+im_{e}rg_{2}-(\widehat{\mathcal{L}}_{-}f_{1}-am_{e}\cos{\vartheta}g_{2})&=&0.
\end{eqnarray*}
Let us now define
\begin{eqnarray*}
&&f_{1}(r,\vartheta)=\gamma_{1}(r)\delta_{1}(\vartheta),\quad f_{2}(r,\vartheta)=\sigma_{2}(r)\tau_{2}(\vartheta),\\
&&g_{1}(r,\vartheta)=\alpha_{1}(r)\beta_{1}(\vartheta), \quad \hspace{-1mm}g_{2}(r,\vartheta)=\epsilon_{2}(r)\mu_{2}(\vartheta).
\end{eqnarray*}
It can be easily seen that the separability conditions for the above system are 
\[ 
\beta_{1}(\vartheta)=\delta_{1}(\vartheta),\quad \epsilon_{2}(r)=\gamma_{1}(r),\quad\mu_{2}(\vartheta)=\tau_{2}(\vartheta),\quad \alpha_{1}(r)=\sigma_{2}(r).
\]
Following Chandrasekhar (1976), we set
\[ 
\gamma_{1}(r)=R_{-}(r),\quad \delta_{1}(\vartheta)=S_{-}(\vartheta),\quad \sigma_{2}(r)=R_{+}(r),\quad \tau_{2}(\vartheta)=S_{+}(\vartheta).
\]
and we get for $\widetilde{\psi}(r,\vartheta)$ the so called Chandrasekhar ansatz
\numparts
\begin{eqnarray} \label{Chandra1}
&&f_{1}(r,\vartheta)=R_{-}(r)S_{-}(\vartheta),\quad f_{2}(r,\vartheta)=R_{+}(r)S_{+}(\vartheta),\\
&&g_{1}(r,\vartheta)=R_{+}(r)S_{-}(\vartheta),\quad g_{2}(r,\vartheta)=R_{-}(r)S_{+}(\vartheta),
\end{eqnarray}
\endnumparts
by means of which the above system of partial differential equations becomes
\numparts
\begin{eqnarray}\label{c1}
(\sqrt{\Delta}\widehat{\mathcal{D}}_{+}R_{+}+im_{e}rR_{-})S_{-}+(\widehat{\mathcal{L}}_{+}S_{+}-am_{e}\cos{\vartheta}S_{-})R_{-}&=&0,\\
(\sqrt{\Delta}\widehat{\mathcal{D}}_{-}R_{-}-im_{e}rR_{+})S_{+}-(\widehat{\mathcal{L}}_{-}S_{-}-am_{e}\cos{\vartheta}S_{+})R_{+}&=&0,\\
(\sqrt{\Delta}\widehat{\mathcal{D}}_{-}R_{-}-im_{e}rR_{+})S_{-}+(\widehat{\mathcal{L}}_{+}S_{+}-am_{e}\cos{\vartheta}S_{-})R_{+}&=&0,\\
(\sqrt{\Delta}\widehat{\mathcal{D}}_{+}R_{+}+im_{e}rR_{-})S_{+}-(\widehat{\mathcal{L}}_{-}S_{-}-am_{e}\cos{\vartheta}S_{+})R_{-}&=&0.
\end{eqnarray}
\endnumparts
Introducing four separation constants $\lambda_{1},\dots,\lambda_{4}$ as follows
\numparts
\begin{eqnarray}\label{b1}
\widehat{\mathcal{L}}_{+}S_{+}-am_{e}\cos{\vartheta}S_{-}&=&-\lambda_{1}S_{-},\\
\widehat{\mathcal{L}}_{-}S_{-}-am_{e}\cos{\vartheta}S_{+}&=&+\lambda_{2}S_{+},\\
\widehat{\mathcal{L}}_{+}S_{+}-am_{e}\cos{\vartheta}S_{-}&=&-\lambda_{3}S_{-},\\
\widehat{\mathcal{L}}_{-}S_{-}-am_{e}\cos{\vartheta}S_{+}&=&+\lambda_{4}S_{+},
\end{eqnarray}
\endnumparts
we obtain from (\ref{c1})-(2.33{\it{d}})
\numparts
\begin{eqnarray}\label{d1}
\sqrt{\Delta}\widehat{\mathcal{D}}_{+}R_{+}+im_{e}rR_{-}&=&\lambda_{1}R_{-},\\
\sqrt{\Delta}\widehat{\mathcal{D}}_{+}R_{+}+im_{e}rR_{-}&=&\lambda_{4}R_{-},\\
\sqrt{\Delta}\widehat{\mathcal{D}}_{-}R_{-}-im_{e}rR_{+}&=&\lambda_{2}R_{+},\\
\sqrt{\Delta}\widehat{\mathcal{D}}_{-}R_{-}-im_{e}rR_{+}&=&\lambda_{3}R_{+}.
\end{eqnarray}
\endnumparts
Clearly, the systems of equations (\ref{b1})-(2.34{\it{d}}) and (\ref{d1})-(2.35{\it{d}}) will be consistent if $\lambda_{1}=\lambda_{2}=\lambda_{3}=\lambda_{4}=:\lambda$. Notice that the decoupled equations give rise to the following two systems of linear first order differential equations for the radial and angular components of the spinor $\widehat{\psi}$ 
\begin{eqnarray} 
&&\left( \begin{array}{cc}
     \sqrt{\Delta}\widehat{\mathcal{D}}_{-}&-im_{e}r-\lambda\\
     im_{e}r-\lambda&\sqrt{\Delta}\widehat{\mathcal{D}}_{+}
           \end{array} \right)\left( \begin{array}{cc}
                                     R_{-} \\
                                     R_{+}
                                     \end{array}\right)=0, \label{radial}\\
&&\left( \begin{array}{cc}
     -\widehat{\mathcal{L}}_{-} & \lambda+am_{e}\cos\theta\\
                \lambda-am_{e}\cos\theta & \widehat{\mathcal{L}}_{+}
           \end{array} \right)\left( \begin{array}{cc}
                                     S_{-} \\
                                     S_{+}
                                     \end{array}\right)=0. \label{angular}
\end{eqnarray} 
For $a=0$ the components $S_{\pm}$ of the angular eigenfunctions can be expressed in terms of spin-weighted spherical harmonics (see for details Newman and Penrose (1966) and Goldberg \etal. (1967)) whereas in the more general case $a\neq 0$ it can be shown that they satisfy a generalized Heun equation (see Batic \etal. (2005)).\\   
Starting from the systems of equations (\ref{b1})-(2.34{\it{d}}) and (\ref{d1})-(2.35{\it{d}}), we can construct the symmetry operator $J$ for the Dirac operator $\mathcal{O}_{D}$ defined in (\ref{uno}). First of all we set $\lambda_{1}=\lambda_{2}=\lambda_{3}=\lambda_{4}=:\lambda$ in (\ref{b1})-(2.34{\it{d}}) and (\ref{d1})-(2.35{\it{d}}). After multiplication of (\ref{b1}) and (\ref{d1}) by $-rR_{-}$ and $ia\cos{\vartheta}S_{-}$, respectively, and summation of the resulting equations we obtain
\begin{equation} \label{h1}
-ia\overline{\widetilde{\rho}}\cos{\vartheta}\sqrt{\Delta}\widehat{\mathcal{D}}_{+}g_{1}+r\overline{\widetilde{\rho}}\widehat{\mathcal{L}}_{+}g_{2}=\lambda f_{1}
\end{equation}
with $\overline{\widetilde{\rho}}=-(r+ia\cos{\vartheta})^{-1}$. Proceeding analogously we obtain from the remaining equations  
\numparts
\begin{eqnarray}\label{h2}
&&-ia\overline{\widetilde{\rho}}\cos{\vartheta}\sqrt{\Delta}\widehat{\mathcal{D}}_{-}g_{2}-r\overline{\widetilde{\rho}}\widehat{\mathcal{L}}_{-}g_{1}=\lambda f_{2},\\
&&+ia\widetilde{\rho}\cos{\vartheta}\sqrt{\Delta}\widehat{\mathcal{D}}_{-}f_{1}+r\widetilde{\rho}\widehat{\mathcal{L}}_{+}f_{2}=\lambda g_{1},\\
&&+ia\widetilde{\rho}\cos{\vartheta}\sqrt{\Delta}\widehat{\mathcal{D}}_{+}f_{2}-r\widetilde{\rho}\widehat{\mathcal{L}}_{-}f_{1}=\lambda g_{2},
\end{eqnarray}
\endnumparts
with $\rho=-(r-ia\cos{\vartheta})^{-1}$. By means of (\ref{h1}) and (\ref{h2}\,\textit{-c}) we can obtain an explicit expression for the matrix operator $\widehat{J}$, namely
\[
\hspace{-1.8cm}\widehat{J}=\left( \begin{array}{cccc}
0&0&-ia\overline{\widetilde{\rho}}\cos{\vartheta}\sqrt{\Delta}\mathcal{D}_{+}&r\overline{\widetilde{\rho}}\mathcal{L}_{+}\\
0&0&-r\overline{\widetilde{\rho}}\mathcal{L}_{-}&-ia\overline{\widetilde{\rho}}\cos{\vartheta}\sqrt{\Delta}\mathcal{D}_{-}\\
ia\widetilde{\rho}\cos{\vartheta}\sqrt{\Delta}\mathcal{D}_{-}&r\widetilde{\rho}\mathcal{L}_{+}&0&0\\
-r\widetilde{\rho}\mathcal{L}_{-}&ia\widetilde{\rho}\cos{\vartheta}\sqrt{\Delta}\mathcal{D}_{+}&0&0
\end{array} \right).
\]
According to Carter and McLenaghan $\widehat{J}$ can be written in a more compact form as follows
\begin{equation} \label{nove}
\widehat{J}=\Gamma^{-1}(\Gamma_{(\vartheta)}\mathcal{W}_{(t,r,\varphi)}-\Gamma_{(r)}\mathcal{W}_{(t,\vartheta,\varphi)}).
\end{equation}
Starting from $\widehat{J}$ the next lemma shows how to construct a new operator $J$ that will commute with the Dirac operator (\ref{uno}).
\begin{lemma}
Let $\widehat{J}$ be defined as in (\ref{nove}). Then in the Kerr-Newman metric the operator $J=S\widehat{J}S^{-1}$ with $S$ defined as in Lemma~\ref{lemma1} is a symmetry matrix operator for the formal Dirac operator $\mathcal{O}_{D}$ introduced in (\ref{uno}), i.e.
\[
[\mathcal{O}_{D},J]=0.
\]
\end{lemma}
\begin{proof}
From (\ref{vdoppio}) we have $\mathcal{O}_{D}=S\Gamma^{-1}\mathcal{W}S^{-1}$ acting on the spinor $\Psi$. Let us consider the operator $J=S\widehat{J}S^{-1}$ with $\widehat{J}$ given by (\ref{nove}). Then we find that
\begin{eqnarray*}
\mathcal{O}_{D}J&=&S\Gamma^{-1}\mathcal{W}\Gamma^{-1}(\Gamma_{(\vartheta)}\mathcal{W}_{(t,r,\varphi)}-\Gamma_{(r)}\mathcal{W}_{(t,\vartheta,\varphi)})S^{-1}\\
J\mathcal{O}_D&=&S\Gamma^{-1}(\Gamma_{(\vartheta)}\mathcal{W}_{(t,r,\varphi)}-\Gamma_{(r)}\mathcal{W}_{(t,\vartheta,\varphi)})\Gamma^{-1}\mathcal{W}S^{-1}
\end{eqnarray*}
and therefore $[\mathcal{O}_{D},J]=S\Gamma^{-1}PS^{-1}$ with
\[
\hspace{-3mm}P:=\mathcal{W}\Gamma^{-1}(\Gamma_{(\vartheta)}\mathcal{W}_{(t,r,\varphi)}-\Gamma_{(r)}\mathcal{W}_{(t,\vartheta,\varphi)})-(\Gamma_{(\vartheta)}\mathcal{W}_{(t,r,\varphi)}-\Gamma_{(r)}\mathcal{W}_{(t,\vartheta,\varphi)})\Gamma^{-1}\mathcal{W}.
\]
Making use of the decomposition $\mathcal{W}=\mathcal{W}_{(r)}+\mathcal{W}_{(\vartheta)}$ and of the commutation relations (\ref{sei}), we obtain
\[
\hspace{-5mm}P=\mathcal{W}_{(t,\vartheta,\varphi)}(\Gamma_{(r)}\Gamma^{-1}+\Gamma^{-1}\Gamma_{(\vartheta)})\mathcal{W}_{(t,r,\varphi)}-\mathcal{W}_{(t,r,\varphi)}(\Gamma^{-1}\Gamma_{(r)}+\Gamma_{(\vartheta)}\Gamma^{-1})\mathcal{W}_{(t,\vartheta,\varphi)}.
\]
Finally, with the help of (\ref{4}) it results that
\[  
P=\mathcal{W}_{(t,\vartheta,\varphi)}\mathcal{W}_{(t,r,\varphi)}-\mathcal{W}_{(t,r,\varphi)}\mathcal{W}_{(t,\vartheta,\varphi)}=0
\]
and this completes the proof.\hspace{5mm}\opensquare
\end{proof}

\section{The physical scalar product}\label{sec:4}
The Dirac equation (\ref{uno}) possesses a conserved current (e.g. Wald) on general curved space-times. As a consequence the total charge outside the black hole
\[
C_{0}=C(t_{0})=\int_{\Sigma_{t_{0}}} (\phi_{A}\overline{\phi}_{A^{'}}+\overline{\chi}_{A}\chi_{A^{'}})T^{AA^{'}}\, dV
\] 
is constant throughout time where $T^{a}$ denotes the future oriented normal vector field to a Cauchy surface $\Sigma_{t_{0}}$, normalized so that $T_{a}T^{a}=2$ and $dV$ is the volume form induced on $\Sigma_{t_{0}}$ by the metric $g$. Here, $a$ denotes an abstract tensor index which has to be understood as an unprimed spinor index $A$ and a primed spinor index $A^{'}$ clumped together, i.e. $a=AA^{'}$. We recall that for a given null tetrad the vector field $l^{a}+n^{a}$ is time-like future oriented since it is the sum of two future-oriented null vectors. This allows us to associate to any null tetrad a preferred time-like future-directed vector field $l^{a}+n^{a}$. Thus, it is natural to impose $T^{a}=l^{a}+n^{a}$ which becomes in the Carter tetrad and in Boyer-Lindquist coordinates
\[
T^{a}\nabla_{a}=\frac{2(r^{2}+a^{2})}{\sqrt{2\Delta\Sigma}}\left(\frac{\partial}{\partial t}+\frac{a}{r^{2}+a^{2}}\frac{\partial}{\partial \varphi}\right).
\]
Taking into account that for a generic null tetrad it was shown by H\"{a}fner and Nicolas (2004) that
\[
T^{AA^{'}}=\left(\begin{array}{cc}
                            n_{a}T^{a}&-\overline{m}_{a}T^{a}\\
                            -m_{a}T^{a}&l_{a}T^{a}
                \end{array}\right)
\]
and employing (\ref{t1})-(2.9{\it{c}}) a simple computation gives that the matrix $T^{AA^{'}}$ is the identity. Therefore the total charge outside the black hole is
\[
C_{0}=\int_{\Sigma_{t_{0}}} (|\phi_{0}|^{2}+|\phi_{1}|^{2}+|\chi_{0^{'}}|^{2}+|\chi_{1^{'}}|^{2})\, dV
\]
which is manifestly a positive quantity. By means of Theorem 13.3 in Oloff the volume form $dV$ induced on $\Sigma_{t_{0}}$ by the Kerr-Newman metric is computed to be
\[
C_{0}=\int_{r_{1}}^{+\infty}\,dr\int_{-1}^{1}\,d(\cos{\vartheta})\int_{0}^{2\pi}\,d\varphi(r^2+a^2)\sqrt{\frac{\widetilde{\Sigma}\Sigma}{\Delta}}|\Psi|^{2}
\]
where the spinor $\Psi$ has been defined in Section~\ref{sec:2}. Let us bring the Dirac equation (\ref{uno}) into the form of a Schr\"{o}dinger equation $i\partial_{t}\Psi=H_{D}\Psi$. Then, the formal Hamiltonian operator $H_{D}$ acting on the spinor $\Psi$ on the hypersurface $\Sigma_{t_{0}}$ is formally self-adjoint with respect to the positive scalar product
\begin{equation}\label{scalar0}
\langle\Psi|\Phi\rangle=\int_{r_{1}}^{+\infty}\,dr\int_{-1}^{1}\,d(\cos{\vartheta})\int_{0}^{2\pi}\,d\varphi(r^2+a^2)\sqrt{\frac{\widetilde{\Sigma}\Sigma}{\Delta}}\overline{\Psi}\Phi.
\end{equation}
If we introduce the tortoise coordinate $u\in\mathbb{R}$ defined by $du/dr=(r^{2}+a^{2})/\Delta$, the scalar product (\ref{scalar0}) can be written in the more compact form
\begin{equation}\label{scalar}
\langle\Psi|\Phi\rangle=\int_{-\infty}^{+\infty}\,du\int_{-1}^{1}\,d(\cos{\vartheta})\int_{0}^{2\pi}\,d\varphi\hspace{1mm}\sqrt{\widetilde{\Sigma}\Sigma\Delta}\hspace{2mm}\overline{\Psi}\Phi.
\end{equation}
Finally, by trasforming the spinors according to $\widetilde{\Psi}=(\widetilde{\Sigma}\Sigma\Delta)^{\frac{1}{4}}\Psi$ and taking into account the corresponding changes in the Hamiltonian $H_{D}$ due to this transformation, we could even work in a standard (i.e. not weighted) $L_{2}$ space.

\section{Reduction to the Schr\"{o}dinger form and self-adjointness of the Dirac Hamiltonian} \label{sec:5}
In order to bring the matrix equation $\mathcal{W}\widehat{\psi}=0$ with $\mathcal{W}$ given by (\ref{unoo}) into the form of a Schr\"{o}dinger equation $i\partial_{t}\widehat{\psi}=h\widehat{\psi}$ we bring the time derivatives in (\ref{unoo}) to the l.h.s. of the equation and find that
\begin{equation} \label{Sch}
-i\mathfrak{T}\partial_{t}\widehat{\psi}=(\mathcal{W}_{(r,\varphi)}+\mathcal{W}_{(\vartheta,\varphi)})\widehat{\psi}
\end{equation} 
with 
\[
\mathfrak{T}=\left( \begin{array}{cccc}
                            0&0&i\frac{r^{2}+a^{2}}{\sqrt{\Delta}}&-a\sin{\vartheta}\\
                            0&0&-a\sin{\vartheta}&-i\frac{r^{2}+a^{2}}{\sqrt{\Delta}}\\
                            -i\frac{r^{2}+a^{2}}{\sqrt{\Delta}}&-a\sin{\vartheta}&0&0\\
                            -a\sin{\vartheta}&i\frac{r^{2}+a^{2}}{\sqrt{\Delta}}&0&0
         \end{array} \right)
\]
and $\mathcal{W}_{(r,\varphi)}$, $\mathcal{W}_{(\vartheta,\varphi)}$ formally given by (\ref{due}) and (\ref{3}), respectively, with $\mathcal{D}_{\pm}$ and $\mathcal{L}_{\pm}$ now replaced by 
\[
D_{\pm}=\frac{\partial}{\partial r}\mp\frac{1}{\Delta}\left(a\frac{\partial}{\partial\varphi}-ieQr\right),\quad L_{\pm}=\frac{\partial}{\partial \vartheta}+\frac{1}{2}\cot{\vartheta}\mp i\csc{\vartheta}\frac{\partial}{\partial\varphi}.
\]
\begin{lemma} \label{polinomio}
 $\mbox{\upshape{det}}(\mathfrak{T})\neq 0$ for every $r>r_{1}>0$ and $\vartheta\in[0,\pi]$.
\end{lemma}
\begin{proof}
A short computation gives $\mbox{\upshape{det}}(\mathfrak{T})=p^{2}(r,\vartheta)/\Delta^{2}$ with $p(r,\vartheta)=(r^{2}+a^{2})^{2}\widetilde{\Sigma}$. Since Lemma~\ref{lemma0} implies that in the non-extreme case (as well as in the extreme case) $\widetilde{\Sigma}$ can never be zero, it follows that $\mathfrak{T}$ is invertible.\hspace{5mm}\opensquare 
\end{proof}
Let $u$ be the tortoise coordinate introduced in Section~\ref{sec:4}. According to Lemma~\ref{polinomio} we can multiply (\ref{Sch}) on both sides by $-\mathfrak{T}^{-1}$ to obtain
\begin{equation} \label{ham3}
h=h_{0}+V(u,\vartheta)
\end{equation}
where the formal differential expression $h_{0}$ is given by 
\begin{equation} \label{hamilt4}
\hspace{-2cm}h_{0}=A(u,\vartheta)\left[\left( \begin{array}{cccc}
                            -\mathcal{E}_{-}&0&0&0\\
                            0&\mathcal{E}_{+}&0&0\\
                            0&0&\mathcal{E}_{+}&0\\
                            0&0&0&-\mathcal{E}_{-}
         \end{array} \right)+\left( \begin{array}{cccc}
                            0&-\mathcal{M}_{+}&0&0\\
                            -\mathcal{M}_{-}&0&0&0\\
                            0&0&0&\mathcal{M}_{+}\\
                            0&0&\mathcal{M}_{-}&0
         \end{array} \right)\right],
\end{equation}
and
\begin{eqnarray}
\hspace{-5mm}&&A(u,\vartheta)=\frac{1}{\widetilde{\Sigma}}\left[1\hspace{-1mm}\rm{I}_{4}-\it{a}\gamma(u)\sin{\vartheta}\left(\begin{array}{cc}
       \sigma_{2}&0\\
       0&-\sigma_{2}
           \end{array} \right)\right], \label{matrixA}\\
\hspace{-5mm}&&V(u,\vartheta)=A(u,\vartheta)\left[m_{e}\gamma(u)\left(\begin{array}{cc}
       0&\frac{1\hspace{-1mm}\rm{I}_{2}}{\widetilde{\rho}} \\
       \frac{1\hspace{-1mm}\rm{I}_{2}}{\overline{\widetilde{\rho}}}&0
           \end{array} \right)-\frac{eQr}{r^2+a^2}1\hspace{-1mm}\rm{I}_{4}\right] \label{potential}
\end{eqnarray}
where $\sigma_{2}$ is a Pauli matrix, $\widetilde{\rho}=-(r-ia\cos{\vartheta})^{-1}$ and
\[
\mathcal{E}_{\pm}=i\frac{\partial}{\partial u}\mp\frac{ia}{r^2+a^2}\frac{\partial}{\partial\varphi},\quad\quad\mathcal{M}_{\pm}=i\gamma(u)L_{\pm},
\]
satisfying $\overline{\mathcal{E}}_{\pm}=-\mathcal{E}_{\pm}$ and $\overline{\mathcal{M}}_{\pm}=-\mathcal{M}_{\mp}$. In the definition of $\mathcal{E}_{\pm}$ we made use of the tortoise coordinate $u\in(-\infty,\infty)$, that we introduced in Section~\ref{sec:4}. Furthermore,  the matrix contained in the square brackets in (\ref{matrixA}) is hermitian.
\begin{remark}
It should be noted that the product of the matrices $B$ and $C$ entering in the equation (2.11) in Finster \etal. (2003), more precisely 
\[
BC=\left(\begin{array}{cc}
-\sigma_{2}&0\\
0&\sigma_{2}
\end{array}\right), 
\]
is in contrast to our result (\ref{matrixA}).
\end{remark}
In view of (\ref{scalar}) in Section~\ref{sec:4} it follows that the formal Hamilton operator $h$ is formally self-adjoint with respect to the positive scalar product 
\begin{equation}\label{scalar1}
\langle\widehat{\psi}|\widehat{\phi}\rangle=\int_{-\infty}^{+\infty}\,du\int_{-1}^{1}\,d(\cos{\vartheta})\int_{0}^{2\pi}\,d\varphi \hspace{1mm}\sqrt{\widetilde{\Sigma}}\hspace{2mm}\overline{\widehat{\psi}}\widehat{\phi}
\end{equation}
where the spinors has been transformed according to $\Psi=S\widehat{\psi}$ with the transformation matrix $S$ given in Lemma~\ref{lemma1} and $\widetilde{\Sigma}$ defined in Section~\ref{sec:2}.
\begin{remark}
As in Finster \etal. (2003) we made use of the Carter tetrad and after a regular and time-independent transformation given in Lemma~\ref{lemma1} we found that the Dirac equation (\ref{unoo}) coincide with their equation (2.1). However the scalar product expressed by (2.15) and (2.16) in Finster \etal. (2003) does not agree with ours (\ref{scalar1}).
\end{remark}
For simplicity in notation we will write $\psi$ instead of $\widehat{\psi}$ when there is no risk of confusion. In what follows we consider the Hilbert space $\mathcal{H}$ consisting of wave functions $\psi:\Omega:=\mathbb{R}\times S^{2}\longrightarrow\mathbb{C}^4$ together with the positive scalar product (\ref{scalar1}). We study now the formal differential expression $h_{0}$ as an operator on
\[
L_{2}(\Omega)^{4}:=L_{2}\left(\Omega,\sqrt{\widetilde{\Sigma}}\hspace{1mm}du\hspace{1mm}d(\cos{\vartheta})\hspace{1mm}d\varphi\right)^{4}.
\] 
To this purpose we introduce the domains $D(H_{0})$ consisting of all $\psi\in L_{2}(\Omega)^{4}$ such that $H_{0}\psi\in L_{2}(\Omega)^{4}$ in the sense of distributions and $D(\dot{T}):=\mathcal{C}_{0}^{\infty}(\Omega)^{4}$ where $H_{0}\psi=h_{0}\psi$ for all $\psi\in D(H_{0})$ and $\dot{T}\psi=h_{0}\psi$ for all $\psi\in D(\dot{T})$.
\begin{lemma}
$\dot{T}$ is essentially self-adjoint and $\overline{\dot{T}}=H_{0}$.
\end{lemma}
\begin{proof}
We show first that $\dot{T}$ is symmetric. Clearly for all $\psi$, $\phi\in D(\dot{T})$ the operator $\dot{T}$ is formally self-adjoint with respect to the scalar product (\ref{scalar1}), i.e. 
\begin{equation} \label{condizione}
\langle \dot{T}\psi|\phi\rangle=\langle \psi|\dot{T}\phi\rangle.
\end{equation}
Since $\dot{T}$ is densely defined and (\ref{condizione}) holds for every $\psi$, $\phi\in D(\dot{T})$, it follows that $\dot{T}$ satisfies the definition of a symmetric operator as given in Birman and Solomjak. The proof of the essentially self-adjointness is the same as that for part (c) of Theorem IX.27 in Reed and Simon vol II.\hspace{5mm}\opensquare
\end{proof}
Notice that, since $\dot{T}$ is essentially self-adjoint, then there is uniquely associated to $\dot{T}$ a self-adjoint operator $\overline{\dot{T}}=\dot{T}^{**}=\dot{T}^{*}=H_{0}$ and we call it the unperturbed Hamiltonian. Let $\dot{\mathcal{S}}$ be the minimal operator constructed from $h_{0}+V$, i.e. $\dot{\mathcal{S}}=\dot{T}+\mathcal{V}=\dot{T}+\mathcal{V}_{0}+\mathcal{V}_{1}$ where $\mathcal{V}$, $\mathcal{V}_{0}$ and $\mathcal{V}_{1}$ denote the maximal multiplication operators by $V$, $V_{0}$ and $V_{1}$ with
\begin{eqnarray*}
&&V(u,\vartheta)=V_{0}(u,\vartheta)+V_{1}(u,\vartheta),\\
&&V_{0}(u,\vartheta)=-\frac{eQr}{r^2+a^2}A(u,\vartheta),\quad V_{1}(u,\vartheta)=m_{e}\gamma(u)A(u,\vartheta)\left(\begin{array}{cc}
       0&\frac{1\hspace{-1mm}\rm{I}_{2}}{\widetilde{\rho}} \\
       \frac{1\hspace{-1mm}\rm{I}_{2}}{\overline{\widetilde{\rho}}}&0
           \end{array} \right).
\end{eqnarray*}

\begin{theorem} \label{esa}
$\dot{\mathcal{S}}$ is essentially self-adjoint and $\overline{\dot{\mathcal{S}}}=H$.
\end{theorem}
\begin{proof}
A simple calculation shows that $V_{0}$, $V_{1}\in L_{\infty}(\Omega)$ since
\[
\|V_{0}\|_{\infty}:=\sup_{x\in\Omega}{\{|V_{0}(x)|\}}<4\frac{|eQ|}{r_{1}},\quad \|V_{1}\|_{\infty}:=\sup_{x\in\Omega}{\{|V_{1}(x)|\}}<4m_{e}.
\]
Hence, $V_{0}$ and $V_{1}$ are bounded and $\dot{\mathcal{S}}$ and $\dot{T}$ have the same domain $C_{0}^{\infty}(\Omega)^{4}$. Let us consider $\dot{\mathcal{S}}$ as obtained from $\dot{T}$ by adding the perturbing term $\mathcal{V}$. In order to show that $\dot{\mathcal{S}}$ is e.s.a., it suffices to prove that $\mathcal{V}$ is relatively bounded with respect to $\dot{T}$ with relative bound $<1$. Let $H$ denotes $\overline{\dot{\mathcal{S}}}$, we show that
\[
H=H_{0}+\mathcal{V},\quad D(H)=D(H_{0})\subset D(\mathcal{V}).
\]
In view of (5.12) in Kato (Ch.V \S 5.3) since $\mathcal{V}_{1}\psi\in L_{2}(\Omega)^{4}$, we have
\[
\|\mathcal{V}_{1}\psi\|_{L_{2}}\leq\|V_{1}\|_{\infty}\|\psi\|_{L_{2}}\leq c\|V_{1}\|_{\infty}(\alpha^{-1/2}\|H_{0}\psi\|_{L_{2}}+\alpha^{3/2}\|\psi\|_{L_{2}})
\]
with some constant $c>0$ and arbitrary $\alpha>0$. On the other hand also $\mathcal{V}_{0}\psi\in L_{2}(\Omega)^{4}$ and $\|\mathcal{V}_{0}\psi\|_{L_{2}}\leq \|V_{0}\|_{\infty}\|\psi\|_{L_{2}}$. Hence, we have $D(\mathcal{V})\supset D(H_{0})\supset D(\dot{T})$ and
\[
\|\mathcal{V}\psi\|_{L_{2}}\leq a\|\psi\|_{L_{2}}+b\|H_{0}\psi\|_{L_{2}}
\]
with $b=c\alpha^{-1/2}\|V_{1}\|_{\infty}$ and $a=c\alpha^{3/2}\|V_{1}\|_{\infty}+\|V_{0}\|_{\infty}$. Since we can choose $\alpha$ arbitrary large, the above inequality shows that $\mathcal{V}$ is $H_{0}$-bounded with relative bound $0$. Hence, Theorem 4.1 in Kato (Ch.V \S 4.1) implies that $\dot{\mathcal{S}}=\dot{T}+\mathcal{V}$ is essentially self-adjoint. Finally, since $H_{0}+\mathcal{V}\supset\dot{\mathcal{S}}$, it follows that $H_{0}+\mathcal{V}=\overline{\dot{\mathcal{S}}}$. Thus, the perturbed operator $H$ has the same domain as the unperturbed operator $H_{0}$.\hspace{5mm}\opensquare
\end{proof}
\begin{remark}
It should be noted that in Finster \etal. (2003) the essentially self-adjointness of $H$ is stated without proof.
\end{remark}

\section{The Dirac propagator in the Kerr-Newman metric}\label{sec:6}
We begin with some preliminary observations. Energy, generalized squared total angular momentum and the z-component of the total angular momentum form a set of commuting observables $\{H,\widehat{J}^{2},\widehat{J}_{z}\}$. Moreover, the angular system (\ref{angular}) can be brought into the so-called Dirac form (see Batic \etal. (2005)), namely
\begin{equation}\label{UU}
\hspace{-1.2cm}\mathcal{U}S:=\left( \begin{array}{cc}
       0&1\\
      -1&0
                     \end{array} \right)\frac{dS}{d\vartheta}+\left( \begin{array}{cc}
       -am_{e}\cos{\vartheta}&-\frac{k+\frac{1}{2}}{\sin{\vartheta}}-a\omega\sin{\vartheta}\\
      -\frac{k+\frac{1}{2}}{\sin{\vartheta}}-a\omega\sin{\vartheta}&am_{e}\cos{\vartheta}
                     \end{array} \right)S=\lambda S
\end{equation}
with $S(\vartheta)=(S_{-}(\vartheta),S_{+}(\vartheta))^{T}$ and $\vartheta\in(0,\pi)$. For the theory of Dirac systems of ordinary differential equations we refer to Weidmann (1987). In $L_{2}(0,\pi)^{2}$ the angular operator $\mathcal{U}$ with domain $D(\mathcal{U})=\mathcal{C}_{0}^{\infty}(0,\pi)^{2}$ is essentially self-adjoint, its spectrum is discrete and it consists of simple eigenvalues, i.e. $\lambda_{j}<\lambda_{j+1}$ for every $j\in\mathbb{Z}\backslash\{0\}$. Moreover, the eigenvalues depend smoothly on $\omega$. For details see Finster \etal. (2000) and Batic \etal. (2005). Furthermore, the functions $e^{i\left(k+\frac{1}{2}\right)\varphi}$ are eigenfunctions of the z-component of the total angular momentum operator $\widehat{J}_{z}$ with eigenvalues $-\left(k+\frac{1}{2}\right)$ with $k\in\mathbb{Z}$. Hence, in the Hilbert space $\mathcal{H}$ we will label the generalized states $\psi$ by $\psi^{kj}_{\omega}$. In what follows, we consider the Dirac operator $H=\overline{\dot{\mathcal{S}}}$ in $\mathcal{H}=L_{2}(\Omega)^{4}$. Finally, notice that Theorem~\ref{esa} implies that the operator $\dot{\mathcal{S}}$ defined on $\mathcal{C}_{0}^{\infty}(\Omega)^{4}$ is essentially self-adjoint and has a unique self-adjoint extension $H=\overline{\dot{\mathcal{S}}}$.\\
In preparation of the proof for the completeness of the Chandrasekhar separation of variables (\ref{Chandra1}\,\textit{-b}) for the Dirac equation in the Kerr-Newman geometry, we show first that the angular eigenfunctions form a complete orthonormal basis in $L_{2}(S^{2})^{2}$.
\begin{lemma}
For every $k\in\mathbb{Z}$, $j\in\mathbb{Z}\backslash\{0\}$ the set $\{Y^{kj}_{\omega}(\vartheta,\varphi)\}$ with 
\begin{equation} \label{ypsilon}
Y^{kj}_{\omega}(\vartheta,\varphi)=\left( \begin{array}{c}
      Y_{\omega,-}^{kj}(\vartheta,\varphi) \\
      Y_{\omega,+}^{kj}(\vartheta,\varphi)
           \end{array} \right)=\frac{1}{\sqrt{2\pi}}\left( \begin{array}{c}
      S_{\omega,-}^{kj}(\vartheta) \\
      S_{\omega,+}^{kj}(\vartheta)
           \end{array} \right)e^{i\left(k+\frac{1}{2}\right)\varphi},
\end{equation}
is a complete orthonormal basis for $L_{2}(S^{2})^{2}$.
\end{lemma}
\begin{proof}
Concerning the orthonormality, we show that for every $k^{'}$, $k\in\mathbb{Z}$ and $j^{'}$, $j\in\mathbb{Z}\backslash\{0\}$ the relation 
\[
\langle Y^{kj}_{\omega}|Y^{k^{'}j^{'}}_{\omega}\rangle_{S^{2}}=\delta_{kk^{'}}\delta_{jj^{'}}
\] 
holds where $\langle\cdot|\cdot\rangle_{S^{2}}$ denotes the scalar product on $S^{2}$. For simplicity in notation we omit to write the subscript $\omega$ attached to the angular eigenfunctions when there is no risk of confusion. With the help of (\ref{ypsilon}) a direct computation gives
\[
\langle Y^{kj}|Y^{k^{'}j^{'}}\rangle_{S^{2}}=\delta_{kk^{'}}\int_{0}^{\pi}d\vartheta\,\sin{\vartheta}~\overline{S}^{kj}(\vartheta)S^{k^{'}j^{'}}(\vartheta).
\]
It suffices now to consider the case $k=k^{'}$. In this case since $S^{kj}$ and $S^{kj^{'}}$ are both eigenfunctions of the same differential operator $\mathcal{U}$ defined in (\ref{UU}), they are orthogonal. Concerning the completeness,  we can apply the projection theorem (e.g. Reed and Simon Theorem II.3 vol I), i.e. we show that in $L_{2}(S^{2})^{2}$ the only element orthogonal to our orthonormal basis is the zero vector. Without loss of generality let us consider
\[
\widetilde{\varphi}=\frac{1}{\sqrt{2\pi}}\left( \begin{array}{c}
\varphi_{1}(\vartheta)\\
\varphi_{2}(\vartheta)
\end{array} \right)e^{i\left(k+\frac{1}{2}\right)\varphi}\quad\mbox{with}\quad\varphi_{i}\in\mathcal{C}_{0}^{\infty}(0,\pi)\quad\mbox{for}\quad i=1,2.
\]
After the variable transformation $\widetilde{x}=\cos{\vartheta}$ we get
\[
\langle\widetilde{\varphi}|Y^{kj}\rangle_{S^{2}}=\int_{-1}^{1}d\widetilde{x}\,\left(\overline{\varphi}_{1}(\widetilde{x})S_{+}(\widetilde{x})+\overline{\varphi}_{2}(\widetilde{x})S_{-}(\widetilde{x})\right),
\]
Moreover, the following estimate holds
\begin{eqnarray*}
\left|\langle\widetilde{\varphi}|Y^{kj}\rangle_{S^{2}}\right|&\leq&\int_{-1}^{1}d\widetilde{x}\,\left(\left|\overline{\varphi}_{1}(\widetilde{x})\right|\left|S_{+}(\widetilde{x})\right|+\left|\overline{\varphi}_{2}(\widetilde{x})\right|\left|S_{-}(\widetilde{x})\right|\right),\\
&\leq&\|\varphi_{1}\|^{2}_{L_{2}}\int_{-1}^{1}d\widetilde{x}\,\left|S_{+}(\widetilde{x})\right|^{2}+\|\varphi_{2}\|^{2}_{L_{2}}\int_{-1}^{1}d\widetilde{x}\,\left|S_{-}(\widetilde{x})\right|^{2},\\
&\leq&d\int_{-1}^{1}d\widetilde{x}\,\left(\left|S_{+}(\widetilde{x})\right|^{2}+\left|S_{-}(\widetilde{x})\right|^{2}\right)=d
\end{eqnarray*}
where in the second line we used H\"{o}lder inequality, in the third line the orthonormality condition for $S_{\pm}(\vartheta)$ and $d:=\max\{\|\varphi_{1}\|^{2}_{L_{2}},\|\varphi_{2}\|^{2}_{L_{2}}\}$. Clearly, $\left|\langle\widetilde{\varphi}|Y^{kj}\rangle_{S^{2}}\right|\leq 0$ if and only if $d=0$. Since $d=0$ implies $\varphi_{1}=\varphi_{2}=0$, $\mathcal{C}_{0}^{\infty}(0,\pi)$ is dense in $L_{2}(0,\pi)$ and the scalar product $\langle\cdot|\cdot\rangle_{S^{2}}$ is continuous, the proof is completed.\hspace{5mm}\opensquare 
\end{proof} 
We state now the theorem on the completeness of the Chandrasekhar ansatz. In what follows let $\sigma(H)\subseteq\mathbb{R}$ denote the spectrum of the self-adjoint Hamiltonian operator $H$.
\begin{theorem} \label{completo}
For every $\psi\in\mathcal{C}_{0}^{\infty}(\Omega)^{4}$
\begin{equation} \label{completeness}
\hspace{-2.5cm}\psi(0,x)=\int_{\sigma(H)}d\omega\,\sum_{j\in\mathbb{Z}\backslash\{0\}}\sum_{k\in\mathbb{Z}}\langle\psi^{kj}_{\omega}|\psi_{\omega}\rangle\psi^{kj}_{\omega}(x),\hspace{-2mm}\quad \psi^{kj}_{\omega}(x)=\left( \begin{array}{c}
               R_{\omega,-}^{kj}(u)Y_{\omega,-}^{kj}(\vartheta,\varphi)\\
               R_{\omega,+}^{kj}(u)Y_{\omega,+}^{kj}(\vartheta,\varphi)\\
               R_{\omega,+}^{kj}(u)Y_{\omega,-}^{kj}(\vartheta,\varphi)\\
               R_{\omega,-}^{kj}(u)Y_{\omega,+}^{kj}(\vartheta,\varphi)
\end{array} \right)
\end{equation}
where the scalar product $\langle\cdot|\cdot\rangle$ is given by (\ref{scalar1}) and $x=(u,\vartheta,\varphi)$.  
\end{theorem}
\begin{proof}
We show first that it is possible to construct isometric operators
\[
\widehat{W}_{k,j}:\mathcal{C}_{0}^{\infty}(\mathbb{R})^{2}\longrightarrow \mathcal{C}_{0}^{\infty}(\Omega)^{4},
\]
such that 
\[
R^{kj}_{\omega}(u)=\left( \begin{array}{c}
               R^{kj}_{\omega,-}(u)\\
               R^{kj}_{\omega,+}(u)
\end{array} \right)\longmapsto\mathcal{A}(u,\vartheta)\left( \begin{array}{c}
               R_{\omega,-}^{kj}(u)Y_{\omega,-}^{kj}(\vartheta,\varphi)\\
               R_{\omega,+}^{kj}(u)Y_{\omega,+}^{kj}(\vartheta,\varphi)\\
               R_{\omega,+}^{kj}(u)Y_{\omega,-}^{kj}(\vartheta,\varphi)\\
               R_{\omega,-}^{kj}(u)Y_{\omega,+}^{kj}(\vartheta,\varphi)
\end{array} \right)
\]
with some function $\mathcal{A}$ to be determined and $Y_{\omega,\pm}^{kj}$ given by (\ref{ypsilon}). Indeed, since the angular eigenfunctions $Y^{kj}_{\omega}$ are normalized, we have 
\[
\hspace{-4mm}\|R^{kj}_{\omega}\|^{2}_{L_{2}(\mathbb{R})^{2}}=\int_{-\infty}^{+\infty}du\int_{-1}^{1}d(\cos{\vartheta})\int_{0}^{2\pi}d\varphi\, \overline{\psi}^{kj}_{\omega}\psi^{kj}_{\omega},\quad\psi^{kj}_{\omega}=\left( \begin{array}{c}
               R_{\omega,-}^{kj}Y_{\omega,-}^{kj}\\
               R_{\omega,+}^{kj}Y_{\omega,+}^{kj}\\
               R_{\omega,+}^{kj}Y_{\omega,-}^{kj}\\
               R_{\omega,-}^{kj}Y_{\omega,+}^{kj}
\end{array} \right).
\]
and by choosing $\mathcal{A}=(\widetilde{\Sigma})^{-\frac{1}{4}}$ it results that $\|R^{kj}_{\omega}\|^{2}_{L_{2}(\mathbb{R})^{2}}=\|\widehat{W}_{k,j}(R^{kj}_{\omega})\|^{2}_{L_{2}(\Omega)^{4}}$. By means of the isometric operators $\widehat{W}_{k,j}$ we can now introduce for every $\omega\in\sigma(H)$ an auxiliary separable Hilbert space $\mathfrak{h}(\omega)$ as follows
\[
\mathfrak{h}(\omega)=\bigoplus_{k,j\in\mathbb{Z}}\mathfrak{h}_{k,j},\quad\mathfrak{h}_{k,j}=\widehat{W}_{k,j}(\mathcal{C}_{0}^{\infty}(\mathbb{R})^{2}).
\]
Moreover, the expansion theorem (e.g. Th.3.7 in Weidmann (1976)) implies that every element $\psi_{\omega}$ in $\mathfrak{h}(\omega)$ can be written as
\begin{equation} \label{opsala}
\psi_{\omega}=\sum_{k,j\in\mathbb{Z}}\langle\psi^{kj}_{\omega}|\psi_{\omega}\rangle\psi^{kj}_{\omega}.
\end{equation}
Notice that since the solutions $R_{\omega,\pm}^{kj}$ of the radial system (\ref{radial}) oscillate asymptotically for $u\to+\infty$ (see Section~\ref{sec:7}), $\mathfrak{h}(\omega)$ is not a subspace of $L_{2}(\Omega)^{4}$ with respect to the spatial measure. Finally, let us recall that the direct integral of Hilbert spaces
\begin{equation} \label{primdir}
\mathfrak{H}=\int_{\sigma(H)}d\omega\,\mathfrak{h}(\omega)
\end{equation}
is defined (see for example Yafaev Ch.1 $\S$5.1) as the Hilbert space of vector valued functions $\psi_{\omega}$ that take values in the auxiliary Hilbert spaces $\mathfrak{h}(\omega)$. By definition our Hilbert space $\mathcal{H}$ will be decomposed into a direct integral (\ref{primdir}) if there is given a unitary mapping $\mathcal{F}$ of $\mathcal{H}$ onto $\mathfrak{H}$, but since the Hamiltonian $H$ is self-adjoint, the spectral representation theorem (e.g. Theorem 7.18 in Weidmann (1976)) implies the existence of such an isomorphism $\mathcal{F}$ and this completes the proof.\hspace{5mm}\opensquare
\end{proof}
Starting from the above result we get now more information about the spectral nature of the Hamiltonian $H$. In what follows we denote by $\sigma_{ac}(H)$ the absolutely continuous part of $\sigma(H)$.
\begin{lemma}
$\sigma(H)$ is purely absolutely continuous and $\sigma(H)=\mathbb{R}$.
\end{lemma}
\begin{proof}
We saw in Theorem~\ref{completo} that every representative element $\psi_{\omega}=(\mathcal{F}\psi)(\omega)$ of the element $\psi\in\mathcal{H}$ in the decomposition (\ref{primdir}) can be written in the form given by (\ref{opsala}). Hence, in order to investigate the spectrum of the Hamiltonian $H$ we are allowed to focus our analysis on the radial system (\ref{radial}) arising from the Dirac equation in the Kerr-Newman geometry when we make the ansatz (\ref{Chandra1}\,\textit{-b}). As a consequence we just need to study the spectrum of the differential operator $\mathcal{R}$ associated to the formal differential system (\ref{radial}) after it is brought into the form of a Dirac system of ordinary differential equations (see Weidmann (1987)). Since this analysis has been already performed in Belgiorno and Martellini (1999) where it was shown that $\sigma(\mathcal{R})=\sigma_{ac}(\mathcal{R})=\mathbb{R}$, we can conlcude from Theorem~\ref{completo} that $\sigma(H)=\sigma_{ac}(H)$.
\end{proof}
\begin{remark}
A straightforward implication of the previous lemma is the absence of pure point spectrum for $H$ i.e. $\sigma_{pp}(H)=\emptyset$ which means that no stationary states around a non-extreme Kerr-Newman black hole can exist. Regarding the non-existence of normalizable time-periodic solutions of the massive Dirac equation in the above mentioned framework see also Theorem 1.1 in Finster \etal. (2000) and proposition 7.1 in H\"{a}fner and Nicolas (2004) for the case of massless Dirac fields in the Kerr geometry. Hence, such black holes cannot form a new kind of atomic system with the charged black hole as its nucleus and around an electronic cloud. This possibility is instead left open in the case of an extreme Kerr-Newman black hole (see Schmid (2004)).
\end{remark}
In the next theorem we give the integral representation of the Dirac propagator in Kerr-Newman black hole manifolds. 
\begin{theorem}\label{Dirprop}
For every $\psi\in\mathcal{C}_{0}^{\infty}(\Omega)^{4}$
\begin{equation} \label{rappresentazione}
\hspace{-0.9cm}\psi(t,x)=e^{itH}\psi(0,x)=\int_{-\infty}^{+\infty}d\omega\,e^{i\omega t}\sum_{j\in\mathbb{Z}\backslash\{0\}}\sum_{k\in\mathbb{Z}}\psi_{\omega}^{kj}(x)\langle\psi^{kj}_{\omega}|\psi_{\omega}\rangle
\end{equation}
with scalar product $\langle\cdot|\cdot\rangle$ defined by (\ref{scalar1}) and $\psi^{kj}_{\omega}(x)$ as in (\ref{completeness}).
\end{theorem}
\begin{proof}
Since $H$ is self-adjoint, Theorem 7.37 in Weidmann (1976) guarantees that $\{U(t)=e^{itH}|t\in\mathbb{R}\}$ is a strongly continuous one-parameter unitary group with $U(t)\psi\in\mathcal{C}^{\infty}_{0}(\Omega)^{4}$ for every $t\in\mathbb{R}$ and every $\psi\in\mathcal{C}^{\infty}_{0}(\Omega)^{4}$. The spectral theorem (e.g. Yafaev Ch.1 $\S$4-5) implies that
\begin{equation} \label{prop}
\langle \phi|e^{itH}\psi\rangle=\int_{-\infty}^{+\infty}d\omega\,e^{i\omega t}\langle \phi_{\omega}|\psi_{\omega}\rangle
\end{equation}
where $\langle \phi_{\omega}|\psi_{\omega}\rangle$ is given in terms of the resolvent of $H$ by the following relation
\begin{equation} \label{resol}
\langle \phi_{\omega}|\psi_{\omega}\rangle=\frac{d}{d\omega}\langle \phi|E(\omega)\psi\rangle=\frac{1}{2\pi i}\lim_{\epsilon\to 0}\langle\phi|[R(\omega-i\epsilon)-R(\omega+i\epsilon)]\psi\rangle
\end{equation}
with $E(\omega)$ the spectral family associated to $H$. In addition Theorem 1.7 in Kato (Ch.10 $\S$1) assures that $\langle \phi_{\omega}|\psi_{\omega}\rangle$ is absolutely continuous in $\omega$, while the unicity of the spectral family $E(\omega)$ and the existence of a spectral family $F(\omega)=\mathcal{F}E(\omega)\mathcal{F}^{-1}$  on $\mathfrak{H}$ follow directly from Theorem 7.15 in Weidmann (1976). Notice that $\mathcal{F}$ is the unitary mapping already defined in Theorem~\ref{completo}. By means of equation (3) in Yafaev Ch.1 $\S$4.2
\[
R(z)=\int_{-\infty}^{+\infty}\,\frac{1}{\widetilde{\omega}-z}dE(\widetilde{\omega}),
\]
we get
\[
\frac{1}{2\pi i}[R(\omega-i\epsilon)-R(\omega+i\epsilon)]=\frac{1}{\pi}\int_{-\infty}^{+\infty}\,\frac{\epsilon}{(\omega-\widetilde{\omega})^{2}+\epsilon^{2}}dE(\widetilde{\omega}).
\]
The above integrand is bounded and integrable. Hence, we may apply Lebesgue dominated convergence theorem to take the limit $\epsilon\to 0$ inside the integral sign. Since
\[
\frac{1}{\pi}\lim_{\epsilon\to 0}\frac{\epsilon}{(\omega-\widetilde{\omega})^{2}+\epsilon^{2}}=\delta(\omega-\widetilde{\omega}),
\] 
it results that
\begin{equation} \label{idi}
\frac{1}{2\pi i}\lim_{\epsilon\to 0}[R(\omega-i\epsilon)-R(\omega+i\epsilon)]=\mbox{Id}.
\end{equation}
Finally, (\ref{rappresentazione}) follows from (\ref{opsala}), (\ref{prop}), (\ref{resol}) and (\ref{idi}).\hspace{5mm}\opensquare
\end{proof}
\begin{remark}
An integral representation for the Dirac propagator in the Kerr-Newman metric can also be found in Finster \etal. (2003). In the derivation of the propagator they do not relate directly the spectral measure to the solutions of the radial and angular systems (\ref{radial}) and (\ref{angular}) but  they consider first the Dirac equation in a finite volume $[u_{1},u_{2}]\times S^{2}$ where certain Dirichlet boundary conditions on the spinors are introduced. With respect to our notation for the radial solutions these boundary conditions reduce to the following boundary conditions for the radial functions
\begin{equation} \label{Felcon}
R_{+}(u_{2})=R_{-}(u_{2}),\quad R_{+}(u_{1})=R_{-}(u_{1}).
\end{equation}
The reason why they do that is that the spectrum of the Hamiltonian is purely discrete in a finite volume  and hence, they are allowed to use the discrete version of the spectral theorem. Then they show that the spectral gaps for the Hamiltonian can be made arbitrary small in the infinite volume limit  $u_{1}\to-\infty$ and $u_{2}\to+\infty$ and this finally leads to the integral representation for the propagator. At this point notice that in such an infinite volume limit (\ref{Felcon}) is in contradiction with the asymptotics of the radial solutions in the next section.
\end{remark}

\section{The asymptotic behaviour of the radial solutions}\label{sec:7}
In this section we give locally uniformly in $\omega$ the asymptotics of the solutions of the radial system   
\begin{equation} \label{radialsol} 
\left( \begin{array}{cc}
     \sqrt{\Delta}\widehat{\mathcal{D}}_{-}&-im_{e}r(u)-\lambda\\
     im_{e}r(u)-\lambda&\sqrt{\Delta}\widehat{\mathcal{D}}_{+}
           \end{array} \right)\left( \begin{array}{cc}
                                     R_{-}(u) \\
                                     R_{+}(u)
                                     \end{array}\right)=0,
\end{equation}
\[
\hspace{-0.9cm}\widehat{\mathcal{D}}_{\pm}=\frac{r^{2}(u)+a^{2}}{\Delta}\frac{d}{du}\mp i\frac{K(u)}{\Delta},\quad K(u)=\omega(r^2(u)+a^2)-eQr(u)+\left(k+\frac{1}{2}\right)a.
\]
for $u\to +\infty$ and for $u\to -\infty$, respectively. Notice that the above system has been obtained from (\ref{radial}) by expressing the derivatives with respect to the radial variable $r$ in terms of the tortoise coordinate $u$ introduced in Section~\ref{sec:4}. Concerning the asymptotics for $u\to +\infty$, we treat the case $|\omega|<m_{e}$ and $|\omega|>m_{e}$, separately. Moreover, for ease in notation we will write $r$ instead of $r(u)$ when no risk of confusion arises.
\begin{lemma}\label{faraway}
Every non trivial solution $R$ of (\ref{radialsol}) for $|\omega|>m_{e}$ behaves asymptotically for $u\to +\infty$ like
\[
R(u)=\left( \begin{array}{cc}
                                     R_{-}(u) \\
                                     R_{+}(u)
                                     \end{array}\right)=\left( \begin{array}{cc}
     \cosh{\Theta}&\sinh{\Theta}\\
     \sinh{\Theta}&\cosh{\Theta}
           \end{array} \right)\left( \begin{array}{cc}
                                     e^{-i\Phi(u)}[f^{\infty}_{-}+\mathcal{O}(u^{-1})] \\
                                     e^{+i\Phi(u)}[f^{\infty}_{+}+\mathcal{O}(u^{-1})]
                                     \end{array}\right)
\]
where $f^{\infty}_{\pm}=c_{\pm}e^{\mp i\Phi(u_{0})}$ for some $u_{0}>0$ and the scalars scalars $c_{-}$, $c_{+}$ are such that $|c_{-}|^2+|c_{+}|^2\neq 0$. Moreover,
\[
\hspace{-0.7cm}\Theta=\frac{1}{4}\log{\left(\frac{\omega+m_{e}}{\omega-m_{e}}\right)},\quad\Phi(u)=\kappa u+\frac{Mm^{2}_{e}-\omega eQ}{\kappa}\log{u},\quad \kappa=\epsilon(\omega)\sqrt{\omega^2-m^2_{e}}.
\]
with sign function $\epsilon(\omega)$ such that $\epsilon(\omega)=+1$ if $\omega>m_{e}$ and $\epsilon(\omega)=-1$ if $\omega<m_{e}$. For $|\omega|<m_{e}$ and $u_{0}>0$ the asymptotic behaviour is
\[
\widehat{R}(u)=\left( \begin{array}{cc}
     \frac{\omega-i\sqrt{m^{2}_{e}-\omega^{2}}}{m_{e}}&\frac{\omega+i\sqrt{m^{2}_{e}-\omega^{2}}}{m_{e}}\\
     1&1
           \end{array} \right)\left( \begin{array}{cc}
                                     e^{-\widehat{\Phi}(u)}[\widehat{f}^{\infty}_{-}+\mathcal{O}(u^{-1})] \\
                                     e^{+\widehat{\Phi}(u)}[\widehat{f}^{\infty}_{+}+\mathcal{O}(u^{-1})]
                                     \end{array}\right)
\] 
with $\widehat{f}^{\infty}_{\mp}=\widehat{c}_{\mp}e^{\pm\widehat{\Phi}(u_{0})}$ for some scalars $\widehat{c}_{-}$ and $\widehat{c}_{+}$ such that $|\widehat{c}_{-}|^2+|\widehat{c}_{+}|^2\neq 0$ and
\[
\widehat{\Phi}(u)=\sqrt{m^{2}_{e}-\omega^{2}} u-\frac{Mm^{2}_{e}-\omega eQ}{\sqrt{m^{2}_{e}-\omega^{2}}}\log{u}.
\]
\end{lemma}
\begin{proof}
Let us bring (\ref{radialsol}) into the following form
\begin{equation} \label{ena}
\frac{dR}{du}=V(u)R(u),\quad V(u)=\left( \begin{array}{cc}
     -i\Omega(u)&\varphi(u)\\
     \overline{\varphi}(u)&i\Omega(u)
           \end{array} \right)
\end{equation}
with
\[
\Omega(u):=\omega-\frac{eQr}{r^2+a^2}+\frac{\left(k+\frac{1}{2}\right)a}{r^2+a^2},\quad\varphi(u):=\frac{\sqrt{\Delta}(\lambda+im_{e}r)}{r^2+a^2}.
\]
Since $V(u)$ converges for $u\to +\infty$ to a constant matrix
\[
V_{\infty}:=\left( \begin{array}{cc}
     -i\omega&im_{e}\\
     -im_{e}&i\omega
           \end{array} \right),
\]
we can decompose $V(u)$ as follows
\[
V(u)=V_{\infty}+\frac{V_{1}}{u}+\mathcal{B}(u)
\]
with
\[
V_{1}=\left( \begin{array}{cc}
     ieQ&\lambda-iMm_{e}\\
     \lambda+iMm_{e}&-ieQ
           \end{array} \right),\quad |\mathcal{B}(u)|\leq \frac{C}{u^2}
\]
for some positive constant $C$. Let us now define a matrix function $\mathfrak{A}(u)$ as follows
\[
\mathfrak{A}(u)=V_{\infty}+\frac{V_{1}}{u},\quad V_{\infty}=\lim_{u\to+\infty}\mathfrak{A}(u).
\]
Notice that
\begin{itemize}
\item
for $u\geq u_{0}>0$ the matrix functions $\mathfrak{A}(u)$ and $\mathcal{B}(u)$ are continuously differentiable and
\[
\int_{u_{0}}^{+\infty}\, du|\mathfrak{A}^{'}(u)|<\infty,\quad \int_{u_{0}}^{+\infty}\, du|\mathcal{B}(r)|<\infty,
\]
\item
for $|\omega|>m_{e}$ the eigenvalues $\lambda_{1}=-i\sqrt{\omega^2-m^2_{e}}$ and $\lambda_{2}=+i\sqrt{\omega^2-m^2_{e}}$ of $V_{\infty}$ are simple.
\end{itemize}
Let $\xi_{i}$ be an eigenvector of $V_{\infty}$ belonging to the eigenvalue $\lambda_{i}$ and let $\lambda_{i}(u)$ denote the eigenvalue of $\mathfrak{A}(u)$ which converges to $\lambda_{i}$ as $u\to+\infty$, then Theorem 11  Ch.IV in Coppel implies that the system (\ref{ena}) has solutions
\[
R_{i}(u)=e^{\int_{u_{0}}^{u}\,ds\lambda_{i}(s)}\left[\xi_{i}+\mathcal{O}\left(\frac{1}{u}\right)\right],\quad i=1,2.
\]
A simple calculation gives
\[
\lambda_{1}(u)=-i\left[\kappa+\frac{Mm^{2}_{e}-\omega eQ}{\kappa}\frac{1}{u}\right]+\mathcal{O}\left(\frac{1}{u^2}\right),\quad \lambda_{2}(u)=-\lambda_{1}(u).
\]
with $\kappa:=+\sqrt{\omega^2-m^2_{e}}$, while the eigenvectors of $V_{\infty}$ are
\[
\xi_{1}=\left(\cosh{\Theta},\sinh{\Theta}\right)^{T}\quad \xi_{2}=\left(\sinh{\Theta},\cosh{\Theta}\right)^{T},\quad \Theta=\frac{1}{4}\log\left(\frac{\omega+m_{e}}{\omega-m_{e}}\right).
\]
Hence, for some scalars $c_{-}$ and $c_{+}$ every non trivial solution $R(u)$, i.e. both $c_{-}$ and $c_{+}$ can not be zero at the same time, has asymptotically for $u\to+\infty$ the form  
\[
R(u)=\mathcal{G}(\omega)\left( \begin{array}{cc}
                                     e^{-i\Phi(u)}[f^{\infty}_{-}+\mathcal{O}(u^{-1})] \\
                                     e^{+i\Phi(u)}[f^{\infty}_{+}+\mathcal{O}(u^{-1})]
                                     \end{array}\right),\quad \mathcal{G}(\omega)=\left(\begin{array}{cc}
     \cosh{\Theta}&\sinh{\Theta}\\
     \sinh{\Theta}&\cosh{\Theta}
           \end{array} \right)
\]
where 
\[
\Phi(u):=\kappa u+\frac{Mm^{2}_{e}-\omega eQ}{\kappa}\log{u}, \quad f_{-}^{\infty}:=c_{-}e^{i\Phi(u_{0})},\quad f_{+}^{\infty}:=c_{+}e^{-i\Phi(u_{0})}.
\]
The case $|\omega|<m_{e}$ can be treated analogously.\hspace{5mm}\opensquare
\end{proof}
\begin{remark}
Unlike Lemma 3.5 in Finster \etal. (2003) our method allows us to compute explicitly the expressions for the terms $f^{\infty}_{\pm}$.
\end{remark}
We give now an alternative proof of Lemma 3.1 in Finster \etal.(2003), concerning the asymptotics of the radial solutions close to the event horizon.
\begin{lemma}\label{eventhorizonasympt}
Every non trivial solution $R$ of (\ref{radialsol}) behaves asymptotically for $u\to -\infty$ like
\[
R(u)= \left( \begin{array}{cc}
                                     e^{-i\Omega_{0}u}[f^{(0)}_{-}+\mathcal{O}(e^{du})] \\
                                     e^{+i\Omega_{0}u}[f^{0}_{+}+\mathcal{O}(e^{du})]
                                     \end{array}\right),
\]
with $f^{(0)}_{\pm}=c_{\pm}^{(0)}e^{\mp i\Omega_{0}u_{1}}$ such that $|f^{(0)}_{-}|^2+|f^{(0)}_{+}|^2\neq 0$ and
\[
\Omega_{0}=\omega+\frac{a\left(k+\frac{1}{2}\right)-eQr_{1}}{r^{2}_{1}+a^2},\quad 0<d=4\kappa_{+},\quad\kappa_{+}=\frac{r_{1}-r_{0}}{2(r_{1}^2+a^2)}
\] 
where $\kappa_{+}$ is the surface gravity at the black hole horizon.
\end{lemma}
\begin{proof}
Let us bring again (\ref{radialsol}) into the form (\ref{ena}). Taking into account the definition of the tortoise coordinate $u$ introduced in Section~\ref{sec:4}, it is easy to verify that
\begin{equation}\label{Tort}
u(r)=\frac{1}{2\kappa_{+}}\log{(r-r_{1})}+\mathcal{O}(r-r_{1}),\quad 0<\kappa_{+}=\frac{r_{1}-r_{0}}{2(r_{1}^2+a^2)}
\end{equation}
where $\kappa_{+}$ denotes the surface gravity at the black hole horizon. Taking into account that the potential $V(u)$ entering in (\ref{ena}) converges for $u\to-\infty$ to a constant matrix
\[
V_{0}=\lim_{u\to-\infty}V(u)=\left( \begin{array}{cc}
     -i\Omega_{0}&0\\
     0&-i\Omega_{0}
           \end{array} \right),\quad \Omega_{0}=\omega+\frac{a\left(k+\frac{1}{2}\right)-eQr_{1}}{r^{2}_{1}+a^2},
\]
(\ref{Tort}) suggests an asymptotic expansion of $V(u)$ in powers of $e^{2\kappa_{+}u}$. A short computation gives
\[
V(u)=V_{0}+\mathcal{O}((e^{2\kappa_{+}u})^{2}).
\] 
and applying Theorem 1 Ch.IV in Coppel it results that for $u\to-\infty$ the system (\ref{ena}) has solutions
\[
\hspace{-2.6cm}R_{-}(u)=e^{-i\Omega_{0}(u-u_{1})}\left( \begin{array}{cc}
                                     1 \\
                                     0
                                     \end{array}\right)+\mathcal{O}((e^{du}),\quad R_{+}(u)=e^{i\Omega_{0}(u-u_{1})}\left( \begin{array}{cc}
                                     0 \\
                                     1
                                     \end{array}\right)+\mathcal{O}((e^{du}),\hspace{-1mm}\quad d=4\kappa_{+} 
\]
for some $u_{1}\in\mathbb{R}$ such that $u\leq u_{1}<0$. Hence, for some scalars $c_{-}^{(0)}$ and $c_{+}^{(0)}$ every non trivial solution $R(u)$ has asymptotically for $u\to-\infty$ the form  
\[
R(u)=\left( \begin{array}{cc}
                                     e^{-i\Omega_{0}u}[f^{(0)}_{-}+\mathcal{O}(e^{du})] \\
                                     e^{+i\Omega_{0}u}[f^{(0)}_{+}+\mathcal{O}(e^{du})]
                                     \end{array}\right),\quad f^{(0)}_{\pm}=c_{\pm}^{(0)}e^{\mp i\Omega_{0}u_{1}}.
\]
This completes the proof.\hspace{5mm}\opensquare
\end{proof}
\begin{remark}
Unlike Finster \etal. (2003) by means of our method we get analytical expressions for the terms $f^{\infty}_{\pm}$. Moreover we can show that their constant $d$ entering in the asymptotics for the radial solutions at the event horizon can be expressed in terms of the surface gravity at the black hole horizon. This implies that $d$ does not depend on $\omega$ since $\kappa_{+}$ depends only on the parameters $a$, $Q$ and $M$. Thus the assertion in the statement of Lemma 3.1 in Finster \etal. (2003) that $d$ can be chosen locally uniformly in $\omega$ is somewhat misleading.
\end{remark}

\section{The Dirac equation in oblate spheroidal coordinates} 
In Section~\ref{sec:2} we saw that by setting $M=Q=0$ the Kerr-Newman metric goes over to the Minkowski metric in oblate spheroidal coordinates (\ref{Mink}) since by means of the coordinate transformation (\ref{OSCco}) the expression (\ref{Mink}) can be brought into the form of the Minkowski metric in cartesian coordinates. Concerning the computation of the Dirac equation in the Minkowski space-time in oblate spheroidal coordinates (OSC), we can again use the Carter tetrad (\ref{t1}\,\textit{-c}) with $\Delta$ replaced now by $r^2+a^2$ and proceed exactly as we did in Section~\ref{sec:2} to end up with
\begin{equation} \label{unoOSC}
\mathcal{O}_{OSC}\Psi_{OSC}=\left(\begin{array}{cccc} 
-m_{e}&0&\alpha_{+}&\beta_{+}\\
0&-m_{e}&\beta_{-}&\alpha_{-}\\
\widetilde{\alpha}_{-}&-\widetilde{\beta}_{+}&-m_{e}&0\\
-\widetilde{\beta}_{-}&\widetilde{\alpha}_{+}&0&-m_{e}
\end{array}\right)\Psi_{OSC}=0.
\end{equation}
with
\begin{eqnarray*}
&&\alpha_{\pm}=\pm i\sqrt{\frac{r^2+a^2}{\Sigma}}\left(\mathcal{D}_{\pm}+f(r,\vartheta)\right),\quad\beta_{\pm}=\frac{i}{\sqrt{\Sigma}}\left(\mathcal{L}_{\pm}+g(r,\vartheta)\right),\\
&&\widetilde{\alpha}_{\pm}=\pm i\sqrt{\frac{r^2+a^2}{\Sigma}}\left(\mathcal{D}_{\pm}+\overline{f}(r,\vartheta)\right),\quad\widetilde{\beta}_{\pm}=\frac{i}{\sqrt{\Sigma}}\left(\mathcal{L}_{\pm}+\overline{g}(r,\vartheta)\right),
\end{eqnarray*}
\[
\hspace{-1.6cm}\widetilde{\mathcal{D}}_{\pm}=\frac{\partial}{\partial r}\mp\left[\frac{\partial}{\partial t}+\frac{a}{r^2+a^2}\frac{\partial}{\partial\varphi}\right],\quad \mathcal{L}_{\pm}=\frac{\partial}{\partial \vartheta}+\frac{1}{2}\cot{\vartheta}\mp i\left(a\sin{\vartheta}\frac{\partial}{\partial t}+\csc{\vartheta}\frac{\partial}{\partial\varphi}\right),
\]
\[
f(r,\vartheta)=\frac{1}{2}\left(\frac{r-M}{r^2+a^2}+\frac{1}{r+ia\cos{\vartheta}}\right),\quad g(r,\vartheta)=-\frac{ia\sin{\vartheta}}{2(r+ia\cos{\vartheta})}.
\]
The Dirac equation (\ref{unoOSC}) can be replaced by a modified but equivalent equation
\[
\mathcal{W}_{OSC}\widehat{\psi}_{OSC}=(\Gamma \widetilde{S}^{-1}\mathcal{O}_{OSC}\widetilde{S})\widehat{\psi}_{OSC}=0.
\]
With the help of Corollary~\ref{corollary1} we can decompose $\mathcal{W}_{OSC}$ as follows
\begin{eqnarray*}
&&\mathcal{W}^{(OSC)}=\mathcal{W}_{(t,r,\varphi)}^{(OSC)}+\mathcal{W}_{(t,\vartheta,\varphi)},\\
&&\mathcal{W}^{(OSC)}_{(t,r,\varphi)}=\left( \begin{array}{cccc}
                            im_{e}r&0&\sqrt{r^2+a^2}\widetilde{\mathcal{D}}_{+}&0\\
                            0&-im_{e}r&0&\sqrt{r^2+a^2}\widetilde{\mathcal{D}}_{-}\\
                            \sqrt{r^2+a^2}\widetilde{\mathcal{D}}_{-}&0&-im_{e}r&0\\
                             0&\sqrt{r^2+a^2}\widetilde{\mathcal{D}}_{+}&0&im_{e}r
                            \end{array} \right)
\end{eqnarray*}
where $\mathcal{W}_{(t,\vartheta,\varphi)}$ is given by (\ref{3}). Notice that the angular operator $\mathcal{W}_{(t,\vartheta,\varphi)}$ for the Dirac equation in the Kerr-Newman metric does not change when we go over to the Minkowski metric in OSC since it does not depend on the black hole parameters $M$ and $Q$. If we now apply the Chandrasekhar ansatz as we did in Section~\ref{sec:3}, the equation $\mathcal{W}^{(OSC)}\widehat{\psi}_{OSC}=0$ separetes into the angular system (\ref{angular}) and into the radial system
\begin{equation}\label{radialOSC}
\left( \begin{array}{cc}
     \sqrt{r^2+a^2}\widehat{\mathcal{D}}_{-}&-im_{e}r-\lambda\\
     im_{e}r-\lambda&\sqrt{r^2+a^2}\widehat{\mathcal{D}}_{+}
           \end{array} \right)\left( \begin{array}{cc}
                                     R_{-}^{OSC} \\
                                     R_{+}^{OSC}
                                     \end{array}\right)=0.
\end{equation}
Concerning the construction of a symmetry operator for (\ref{unoOSC}) we can proceed as in Section~\ref{sec:3}. Analogously to the procedure in Section~\ref{sec:5} we bring $\mathcal{W}^{(OSC)}\widehat{\psi}_{OSC}=0$ into the form of a Schr\"{o}dinger equation $i\partial_{t}\widehat{\psi}_{OSC}=H_{OSC}\widehat{\psi}_{OSC}$. Taking into account Section~\ref{sec:4} and that the Minkowski space-time in OSC can be continued analytically through the disc $\mathcal{R}$ defined in Corollary~\ref{corollary1}, it can be verified that the Hamiltonian $H_{OSC}$ is formally self-adjoint with respect to the positive scalar product
\begin{equation}\label{scalar1OSC}
\hspace{-1cm}\langle\widehat{\psi}_{OSC}|\widehat{\phi}_{OSC}\rangle_{OSC}=\int_{-\infty}^{+\infty}\,dr\int_{-1}^{1}\,d(\cos{\vartheta})\int_{0}^{2\pi}\,d\varphi \hspace{1mm}\sqrt{\frac{\Sigma}{r^2+a^2}}\hspace{2mm}\overline{\widehat{\psi}}_{OSC}\widehat{\phi}_{OSC}.
\end{equation}
In what follows for simplicity in notation we write $\psi$ instead of $\widehat{\psi}_{OSC}$. We consider now the free Dirac Hamiltonian $H_{OSC}$ in the Hilbert space 
\[
\widetilde{\mathcal{H}}=\widetilde{L}_{2}(\Omega)^{4}:=L_{2}\left(\Omega,\sqrt{\frac{\Sigma}{r^2+a^2}}\hspace{1mm}dr\hspace{1mm}d(\cos{\vartheta})\hspace{1mm}d\varphi\right)^{4},\quad\Omega=\mathbb{R}\times S^{2}.
\]
From Theorem 1.1 in Thaller the operator $H_{OSC}$ defined on
\[
\widetilde{\mathcal{C}}_{0}^{\infty}(\Omega)^{4}=\mathcal{C}_{0}^{\infty}\left(\Omega,\sqrt{\frac{\Sigma}{r^2+a^2}}\hspace{1mm}dr\hspace{1mm}d(\cos{\vartheta})\hspace{1mm}d\varphi\right)^{4}
\]
 is essentially self-adjoint and has a unique self-adjoint extension on the Sobolev space 
\[
\widetilde{W}^{1,2}(\Omega)^{4}=W^{1,2}\left(\Omega,\sqrt{\frac{\Sigma}{r^2+a^2}}\hspace{1mm}dr\hspace{1mm}d(\cos{\vartheta})\hspace{1mm}d\varphi\right)^{4}.
\]
In addition, the spectrum of $H_{OSC}$ denoted by $\sigma_{OSC}$, is purely absolutely continuous and given by $\sigma_{OSC}:=(-\infty,m_{e}]\cup[m_{e},+\infty)$. We state now the following results without proof since the proofs of Lemma~\ref{lemma1OSC}-3 are analogous, respectively, to the proofs of Theorem~\ref{completo}, Theorem~\ref{Dirprop} and Lemma~\ref{faraway}. 
\begin{lemma} \label{lemma1OSC}
For every $\psi\in\widetilde{\mathcal{C}}_{0}^{\infty}(\Omega)^{4}$
\[
\hspace{-1.2cm}\psi(0,x)=\int_{\sigma_{OSC}}d\omega\,\sum_{j\in\mathbb{Z}\backslash\{0\}}\sum_{k\in\mathbb{Z}}\psi^{kj}_{\omega}(x)\langle\psi^{kj}_{\omega}|\psi_{\omega}\rangle,\quad \psi^{kj}_{\omega}(x)=\left( \begin{array}{c}
               R_{\omega,-}^{kj}(r)Y_{\omega,-}^{kj}(\vartheta,\varphi)\\
               R_{\omega,+}^{kj}(r)Y_{\omega,+}^{kj}(\vartheta,\varphi)\\
               R_{\omega,+}^{kj}(r)Y_{\omega,-}^{kj}(\vartheta,\varphi)\\
               R_{\omega,-}^{kj}(r)Y_{\omega,+}^{kj}(\vartheta,\varphi)
\end{array} \right)
\]
where $x=(r,\vartheta,\varphi)$, the scalar product $\langle\cdot|\cdot\rangle$ is given by (\ref{scalar1OSC}) and $R_{\omega,\pm}^{kj}$ are the radial solutions of the system (\ref{radialOSC}).  
\end{lemma}
\begin{lemma}\label{2OSC}
For every $\psi\in\widetilde{\mathcal{C}}_{0}^{\infty}(\Omega)^{4}$
\[
\psi(t,x)=e^{itH_{OSC}}\psi(0,x)=\int_{\sigma_{OSC}}d\omega\,e^{i\omega t}\sum_{j\in\mathbb{Z}\backslash\{0\}}\sum_{k\in\mathbb{Z}}\psi_{\omega}^{kj}(x)\langle\psi^{kj}_{\omega}|\psi_{\omega}\rangle,
\]
with scalar product $\langle\cdot|\cdot\rangle$ defined by (\ref{scalar1OSC}) and $\psi^{kj}_{\omega}(x)$ as in Lemma~\ref{lemma1OSC}.
\end{lemma}
\begin{lemma}\label{lemma3OSC}
Every non trivial solution $R$ of (\ref{radialOSC}) behaves asymptotically for $r\to +\infty$ like
\[
R(r)=\left( \begin{array}{cc}
                                     R_{-}(r) \\
                                     R_{+}(r)
                                     \end{array}\right)=\left( \begin{array}{cc}
     \cosh{\Theta}&\sinh{\Theta}\\
     \sinh{\Theta}&\cosh{\Theta}
           \end{array} \right)\left( \begin{array}{cc}
                                     e^{-i\kappa r}[\widetilde{f}^{\infty}_{-}+\mathcal{O}(r^{-1})] \\
                                     e^{+i\kappa r}[\widetilde{f}^{\infty}_{+}+\mathcal{O}(r^{-1})]
                                     \end{array}\right)
\]
where $\widetilde{f}^{\infty}_{\pm}=\widetilde{c}_{\pm}e^{\mp i\kappa \widetilde{r}_{0}}$ for some $\widetilde{r}_{0}>0$ and scalars $\widetilde{c}_{-}$ and $\widetilde{c}_{+}$ such that $|\widetilde{c}_{-}|^2+|\widetilde{c}_{+}|^2\neq 0$. Moreover, $\kappa$ and $\Theta$ are defined as in Lemma~\ref{faraway}.
\end{lemma}

\ack
The research of D.B. was supported by the EU grant HPRN-CT-2002-00277. The authors thanks Dr. Monika Winklmeier, Universit\"{a}t Bremen, Germany, for fruitful discussions.

\section*{References}
\begin{harvard}
\item[] 
Batic D, Schmid H and Winklmeier M 2005 {\JMP} {\bf 46} 012504-39 
\item[]
Belgiorno F and Martellini M 1999 {\it Phys. Lett.} B {\bf 453} 17--22
\item[]
Birman M S and Solomjak M Z 1987 {\it Spectral Theory of Self-Adjoint Operators in Hilbert Space} (Dordrecht: D.Ridel Publishing Company)
\item[]
Boyer R H and Lindquist R W 1967 {\JMP} {\bf 8} 265-81
\item[]
Carter B and McLenaghan R G 1979 {\it Phys. Rev.} D {\bf 19}, 1093-97
\item[]
Carter B 1987 {\it Gravitation in Astrophysics (NATO ASI series} B vol 156) ed B Carter and J B Hartle (New York and London: Plenum Press) pp~114-17
\item[]
Chandrasekhar S  1976 {\it Proc. Roy. Soc. London} A {\bf 349} 571-75
\item[]
Chandrasekhar S 1992 {\it The Mathematical Theory of Black Holes} (Oxford: Clarendon Press)
\item[]
Coppel W A 1965 {\it Stability and Asymptotic Behavior of Differential Equations} (Boston: D C Heath and Co)
\item[]
Daud$\acute{\mbox{e}}$ T 2004 {\it Sur la th$\acute{\mbox{e}}$orie de la diffusion pour des champs de Dirac dans divers espaces-temps de la relativit$\acute{\mbox{e}}$ g$\acute{\mbox{e}}$n$\acute{\mbox{e}}$rale} PhD  Thesis, University of Bordoux I
\item[]
Finster F, Kamran N, Smoller J and Yau S T 2000 {\it Commun. Pure Appl. Math.} {\bf 53} 902-29
\item[]
Finster F, Kamran N, Smoller J and Yau S T 2003 {\it Adv. Theor. Math. Phys.} {\bf 7} 25-52
\item[]
Flammer C 1957 {\it Spheroidal Wave Functions} (Stanford: Stanford University Press)
\item[]
Floyd R 1973 {\it The Dynamics of Kerr Fields} PhD Thesis, University of London
\item[]
Goldberg J and Sachs R 1962 {\it Acta Phys. Polon.} {\bf 22} Supp.13 12-23
\item[]
Goldberg J N, Macfarlane A J, Newman E T and Sudarsan E C G 1967 {\JMP} {\bf 8} 2155-61
\item[]
H\"{a}fner D 2003  {\it Dissertationes Math.} {\bf 421} 1-102
\item[]
H\"{a}fner D and Nicolas J P 2004 {\it Rev. Math. Phys.} {\bf 16} No.1 29-123
\item[]
Kato T 1966 {\it Perturbation Theory for linear operators (Die Grundlehren der mathematischen Wissenschaften} Band 132) (Berlin, Heidelberg, New York: Springer Verlag)
\item[]
Kinnersley W 1969 {\JMP} {\bf 10} 1195-1203
\item[]
Mourre E 1981 {\it Comm. Math. Phys} {\bf 78} 391-408
\item[]
Newman E T and Penrose R 1966 {\JMP} {\bf 7} 863-70
\item[]
O'Neill B 1995 {\it The Geometry of Kerr Black Holes} (Wellesley, MA: A. K. Peters)
\item[]
Oloff R. 2002 {\it Geometrie der Raumzeit} (Braunschweig, Wiesbaden: Vieweg)
\item[]
Page D 1976 {\it Phys. Rev.} D {\bf 14} 1509-10
\item[]
Penrose R 1973 {\it Ann. N. Y. Acad. Sci.} {\bf 224} 125-34
\item[]
Penrose R and Rindler W 1986 {\it Spinors and Space-Time} vol 1 (Cambridge: Cambridge University Press)
\item[]
Reed M and Simon B 1975 {\it Methods of Modern Mathematical Physics} vol II (New York: Academic Press)
\item[]
Reed M and Simon B 1980 {\it Methods of Modern Mathematical Physics} vol I (New York: Academic Press)
\item[]
Schmid H 2004 {\it Math. Nach.} {\bf 274} 117-29
\item[]
Thaller B 1992 {\it The Dirac Equation (Texts and Monographs in Physics)} (Berlin, Heidelberg, New York: Springer-Verlag)
\item[]
Wald R M 1984 {\it General Relativity} (Chicago: The University of Chicago Press)
\item[]
Walker M and Penrose R 1970 {\it Commun. Math. Phys.} {\bf 18} 265-74 
\item[]
Weidmann J 1976 {\it Lineare Operatoren in Hilbertr\"{a}umen} (Stuttgart: B G Teubner)
\item[]
Weidmann J 1987 {\it Spectral Theory of Ordinary Differential Operators (Lecture Notes in Mathematics} vol 1258) (Berlin, Heidelberg, New York: Springer-Verlag)
\item[]
Yafaev D R 1992 {\it Mathematical Scattering Theory (Translations of Mathematical Monographs} vol 105) (Providence RI: American Mathematical Society)

\end{harvard}

\end{document}